\documentclass[useAMS]{mn2e}
\usepackage{amssymb} 
\usepackage{amsmath}
\usepackage{graphicx,color,layout}
\title[Resistive and magnetized ADAFs with outflow]
{\textbf{Viscous and resistive
    accretion flows with radially self-similar and outflows}}
\author[K. Faghei \& A. Mollatayefeh]{Kazem Faghei\thanks{E-mail: kfaghei@du.ac.ir} 
 and Azam Mollatayefeh \thanks{E-mail: azam.molatayefe@yahoo.com} \\
School of Physics, Damghan University, Damghan, Iran\\
}
\begin{document}



\maketitle

\label{firstpage}

\begin{abstract}
The existence of outflow in accretion flows is confirmed by observations and magnetohydrodynamics (MHD) simulations. In this paper, we study outflows of accretion flows in the presence of resistivity and toroidal magnetic field. The mechanism of energy dissipation in the flow is assumed to be the viscosity and the magnetic diffusivity due to turbulence in the accretion  flow. It is also assumed that the magnetic diffusivity and the kinematic viscosity are not constant and vary by position and $\alpha$-prescription is used for them. The influence of outflow emanating from accretion disc is considered as a sink for mass, angular momentum and energy. The self-similar method is used to solve the integrated equations that govern the behavior of the accretion flow in the presence of outflow. The solutions represent the disc which rotates  faster and becomes cooler for stronger outflows. Moreover, by adding the magnetic diffusivity, the surface density and rotational velocity decrease, while the radial velocity and temperature increase. The study of present model with the magnitude of magnetic field implies that the disc rotates and accretes faster and becomes hotter, while the surface density decreases. The disc thickness increases by adding the magnetic field or resistivity, while it becomes thinner for more losses of mass and energy due to the outflows. 
\end{abstract}


\section{Introduction}
Accretion disc is an important physical object in astrophysics. 
The standard model of thin accretion disc is widely regarded as a successful model for explaining the 
observational features in active galactic nuclei (AGN) and X-ray binaries (Shakura \& Sunyaev 1973). 
However, the standard disk model cannot explain the spectral energy distributions (SED) of many sources, 
such as Sgr A$^*$. 
To understand such systems, the model of advection dominated accretion flow (ADAF) is introduced (Ichimaru 1977; Narayan \& Yi 1994).
In this accretion flows, the energy released due to dissipation processes 
is retained in the fluid rather than being radiated away. 
The models of ADAF place an intermediate position between the
 spherically symmetric accretion flow of  non-rotating fluid (Bondi 1952) and the cool, thin disc 
of classical accretion disc theory (e. g. Pringle 1981).

The observational evidence of accretion flows imply that outflow is an important property in these systems. 
For example, comparison of the accretion luminosities of neutron stars and white dwarfs in quiescence with similar binary companions
implies that the accretion rate onto white dwarfs is larger by three orders of magnitude than that on the surface of neutron star
(Loeb et al. 2001). This indicates significant outflows in accretion flows. Moreover, outflows seem to be common in 
 the nuclei of galaxies. Marrone et al. (2006) suggested the accretion rate of Sgr A$^*$ at small radii,  much smaller than the Bondi radius, 
must be low, below $10^{-7} M_\odot yr^{-1}$. While, Baganoff et al. (2003) estimated the hot plasma surrounding Sgr A$^*$ 
should supply the accretion rate of $10^{-6} M_\odot yr^{-1}$ at the Bondi radius. This significant  difference between the inner and 
outer mass accretion rates indicates mass loss from the accretion flow due to the outflow  (Kawabata \& Mineshige 2009). Thus, 
the accretion discs in the presence of outflow have been studied by several  authors (Knigge 1999; Fukue 2002, 2004; Shadmehri 2008;
 Xie \& Yuan 2008; Kawabata \& Mineshige 2009; Bu et al. 2009; Li \& Cao 2009; Abbassi et al. 2010).

Knigge (1999) derived the radial distribution of dissipation rate and effective temperature across a Keplerian, steady-state, mass losing
 accretion disc, using a simple, parametric approach that is sufficiently general to be applicable to many types of dynamical disc-wind models.
 Fukue (2002) examined a hydrodynamical wind, which emanates from an accretion disc and is driven by thermal and radiation pressure, under an
 one-dimensional approximation along supposed streamlines. In another study, Fukue (2004) studied a supercritical accretion regime, where
 the mass accretion rate was regulated just at the critical rate with the help of wind mass-loss. He derived a critical radius that outside
 of it, the disc is in a radiation-pressure dominated standard state, while inside this radius the disc is in a critical state, where the
 excess mass is expelled by wind, and the accretion rate is kept being just at the critical rate at any radius inside of critical radius. 
Shadmehri (2008) studied the effects of thermal conduction and outflows on ADAFs. He found that in comparison to accretion
flows without winds, the disc which rotates faster and
becomes cooler because of the angular momentum and energy flux which are taking away by the winds.
Xie \&
 Yuan (2008), based on one-and-a-half-dimensional description of the accretion flow, considered the interchange of mass, radial and azimuthal
 momentum, and the energy between the outflow and inflow. Bu et al. (2009) presented the self-similar solutions for ADAFs with outflows and
 ordered magnetic fields. They assumed the magnetic field has a strong toroidal component and a vertical component in addition to a stochastic
 component. They found that the dynamical properties of ADAFs can be significantly changed in the presence of ordered magnetic fields and
 outflows. Abbassi et al. (2010) examined the effects of a hydrodynamical wind on ADAFs in the presence of a toroidal magnetic field under a 
self-similar treatment. Their results implied that in the presence of the wind, the disc temperature decreases due to energy flux, which is taken
 away by winds and the accretion velocity enhances.      

The importance of magnetic diffusivity has been studied in several accreting systems, 
such as the protostellar discs (Stone et al. 2000; Fleming \& Stone 2003), discs in dwarf nova systems (Gammie \& Menou 1998), 
the discs around black holes (Kudoh \& Kaburaki 1996), and Galactic centre
 (Melia \& Kowalenkov 2001; Kaburaki et al. 2010). Moreover, two and three-dimensional MHD simulations  
have shown that resistive dissipation is one of the crucial processes that determines the saturation amplitude of the magnetorotational instability
 (MRI). As, the linear growth rate of MRI can be reduced significantly due to the suppression by ohmic dissipation (Sano et al. 1998;
 Fleming et al. 2000; Masada \& Sano 2008). Moreover, from comparison of ideal and resistive MHD simulations, it seems the magnetic
 diffusivity may play an important role in astrophysical outflows (Fendt \& \v{C}emelji\'{c} 2002; \v{C}emelji\'{c} et al. 2008).

As mentioned, semi-analytical studies of ADAFs with outflow are typically related to systems without magnetic diffusivity.
While, non-ideal
 MHD simulations imply that the resistivity plays an important role on outflows
 (e. g. Fendt \& \v{C}emelji\'{c} 2002; \v{C}emelji\'{c} et al. 2008).  
Akizuki \& Fukue (2006) proposed a self-similar solution for ADAFs with a highly ionized gas. 
Thus, they assumed that the plasma resistivity is
 zero, and only viscosity is due to turbulence and dissipation in the disc. Also, they ignored the effects of outflow/wind in their
 model.
In this paper,  we want to explore the effects of the resistivity on magnetized ADAFs in the presence of outflows. Thus, we adopt the presented solutions by
 Knigge (1999), Akizuki \& Fukue (2006), and Shadmehri (2008). 
The paper is organized as follow. In section 2, the basic equations of constructing a model for ADAF in the presence of toroidal magnetic
 field, resistivity and outflows will be defined. In section 3, self-similar method for solving equations, which govern the behavior of the
 accreting gas will be used. Results of the present model are brought in sections 4 and 5. The summary of the model will appear in section 6.

\section{Basic Equations}
We consider a non-ideal magnetohydrodynamics of steady, axisymmetric, viscous, accreting and rotating flow in presence of a
 purely toroidal magnetic field. We use a cylindrical coordinate ($r$,  $\phi$, $z$) centred on the accreting object. We ignore the
 general-relativistic effects and use Newtonian gravity. Under these assumptions, the continuity equation is 
\begin{equation}
 \frac{\partial}{\partial r}(r \Sigma v_r)+\frac{1}{2 \pi}\frac{\partial \dot{M}_w}{\partial r}=0,
\end{equation}
where  $v_r$  is the radial velocity, $\Sigma=2 \rho H$ is the surface density of the disc, 
 $\rho$ and $H$ being the mid-plane density and half-thickness of the disc, respectively, and $\dot{M}_w$ is the mass loss rate by
 outflow/wind. The half-thickness of the disc is given by $H=c_s \sqrt{1+\Pi}/\Omega_K $, 
where $\Omega_K$ is the Keplerian angular
 velocity, $c_s$ is the sound speed, $\Pi$ is defined below in equation (9). It will be reduced to its traditional form of $H=c_s/\Omega_K$ in
 absence of the toroidal component of magnetic field ($B_\varphi$; equation 8). The sound speed is defined 
as $c_s=(p_{gas}/\rho)^{1/2}$, where $p_{gas}$ is the gas pressure. The cumulative mass-loss rate from the disc can be written as 
\begin{equation}
 \dot{M}_w(r)=\int^{r}_{{r_{in}}} 4\, \pi\, r^{\prime} \, \dot{m}_w(r^\prime)\,  dr^\prime, 
\end{equation}
where $r_{in}$ denotes the radius at the inner edge of the disc and $\dot{m}_w(r)$ is the mass-loss rate per unit area from each disc face.
 Since the mass accretion rate is $\dot{M}=-2\pi r \Sigma v_r$, from the equations (1) and (2), we can write
\begin{equation}
 \frac{\partial \dot{M}}{\partial r}=\frac{\partial \dot{M}_w}{\partial r}.
\end{equation}
Above equation implies that the mass accretion rate varies by radius due to outflow. Thus, we exploit a power-law
 dependence for mass accretion rate as follows (e.g. Blandford \& Begelman 1999)  
\begin{equation}
 \dot{M}(r)=\dot{M}(R) \left(\frac{r}{R}\right)^s,
\end{equation}
where  $R$ is the radius at the outer edge of the disc, $\dot{M}(R)$ is the mass accretion rate at $R$, and $s$ is a free parameter, which
 for a disc without outflow/wind, $s=0$ and in the presence of the outflow/wind, $s > 0$ (e.g. Fukue 2004). The observed broad-band spectra of
 Sgr A$^*$ and soft X-ray transients can also be fitted by ADAF models with moderate outflows, $s \sim 0.3$ -- $0.4$, if the direct heating of 
electrons in ADAFs is efficient (Quataert \& Narayan 1999; Yuan et al. 2003). Equations (1)-(4) imply that
 \begin{equation}
 \dot{m}_w(r)=s\, \frac{\dot{M}(R)}{4 \pi R^2}\left(\frac{r}{R}  \right)^{s-2}.
\end{equation}
The radial equation of momentum is
\begin{equation}
 v_r \frac{d v_r}{d r}=r\,(\Omega^2-\Omega_K^2)-\frac{1}{\Sigma}\frac{d}{d r}(\Sigma c_s^2)-\frac{c_A^2}{r}-\frac{1}{2\Sigma}\frac{d}{d r}(\Sigma c_A^2)
\end{equation}
where $\Omega$ is the angular velocity of the flow and $c_A$ is Alfven speed, which is defined as 
$c_A^2 \equiv B_{\varphi}^2/(4\pi\rho)=2 p_{mag}/\rho$, $p_{mag}$ being the magnetic pressure.

The angular momentum transfer equation  with consideration of the outflow/wind can be written as (e.g. Knigge 1999)
\begin{equation}
 \Sigma v_r \frac{d}{d r} (r^2 \Omega)=\frac{1}{r}\frac{d}{d r}\left(r^3 \nu \Sigma \frac{d \Omega}{d r} \right)-\frac{(l r)^2 \Omega}{2 \pi r} \frac{d \dot{M}_w}{d r}
\end{equation}
where the two terms on the right-hand side of above equation describe the effects of viscous torques due to shear ($\nu$, which is the effective
 kinematic viscosity) and the angular momentum sink provided by the outflow. Here, it will be assumed that matter outflowed at radius $r$ on the
 disc carries away specific angular momentum $(l r)^2 \Omega$. Thus, $l=0$ corresponds to a non-rotating disc wind and $l=1$ to outflowing
 material that carries away the specific angular momentum it had at the point of outflow (Knigge 1999).

The hydrostatic balance in the vertical direction is integrated to
\begin{equation}
\frac{G M}{r^3} H^2=c_s^2 \left[ 1+\frac{1}{2} \left( \frac{c_A}{c_s}\right)^2 \right]=(1+\Pi) c_s^2.
\end{equation}
Here, we introduce the parameter $\Pi$ by
\begin{equation}
\Pi=\frac{p_{mag}}{p_{gas}}=\frac{1}{2} \left(\frac{c_A}{c_s}\right)^2,
\end{equation}
which is the degree of magnetic pressure to the gas pressure. Since we will apply a steady self-similar method to solve system equation, this
 parameter will be constant throughout the disc. Really, this parameter is a function of position and time (Machida et al. 2006;
 Oda et al. 2007; Khesali \& Faghei 2008, 2009). Studies of hot accretion flows represent the typical value of $\Pi$ lies in the
 range $0.01$-$1$ (De Villiers et al. 2003; Beckwith et al. 2008), but here we also consider the magnetically dominated case ($\Pi > 1$).
 Because, MHD simulation by Machida et al. (2006) shows as thermal instability grows in an accretion flow, the magnetic pressure exceeds
 the gas pressure due to the disc shrink in the vertical direction and conservation of the toroidal magnetic flux. This will result in large
 $\Pi$ and forms a magnetically dominated accretion flow (Oda et al. 2007).

We assume both of the viscosity and the diffusivity are due to turbulence in the disc, so that it is reasonable to use these parameters in
 analogy to the $\alpha$-prescription of Shakura \& Sunyaev (1973) for the turbulent
\begin{equation}
 \nu=P_m \eta=\alpha c_s H,
\end{equation}
where $P_m$ is the magnetic Prandtl number of the turbulence assumed a constant of order of unity, $\eta$ is the magnetic diffusivity, and
 $\alpha$ is a free parameter less than unity.

Here, we can write the energy equation considering energy balance in the system. We will assume the energy released due to viscous and resistive
 dissipations can be balanced by the advection cooling and energy loss of outflow (e.g. Shadmehri 2008). Thus,
\begin{eqnarray}
 \nonumber\frac{\Sigma v_r}{ \gamma-1}\frac{d c_s^2}{d r}-2 H v_r c_s^2 \frac{d \rho}{d r}= \\ \frac{\alpha \sqrt{1+\Pi} f  c_s^2}{\Omega_K} 
\nonumber \left[
   \Sigma r^2  \left(\frac{d \Omega}{d r} \right)^2 
+ \frac{H}{2 \pi P_m}\left( \frac{1}{r}  \frac{d}{d r}(r B_\varphi) \right)^2
\right]\\
-\frac{1}{2} \xi \dot{m}_w(r) v_K^2(r),
\end{eqnarray}
where $f$ is  a constant less than unity and is called advection parameter. The parameter $f$ measures how much the flow is
 advection-dominated (Narayan \& Yi 1994). The first two terms on the right-hand side of above equation represent the  energy generated due to
 viscous and resistive dissipation, respectively. The resistive dissipation is derived by $(4\pi/c^2)\eta \textbf{J}^2$, where
 $\textbf{J}=(c/4\pi)\nabla \times \textbf{B}$ is the current density. Moreover, 
the last term on the right-hand side of energy equation is the
 energy loss due to wind or outflow. 
Depending on the energy loss mechanism, dimensionless parameter $\xi$ may change. We consider
 it as a free parameter of our model so that larger $\xi$ corresponds to more energy extraction from the disc due to the outflows
 (Knigge 1999; Shadmehri 2008).

The  creation/escape rate of the magnetic field can be described by dynamo and diffusion. We define the advection rate of the toroidal
 magnetic field as (Oda et al. 2007)
\begin{equation}
 \dot{\Phi}=\int v_r B_\varphi dz,
\end{equation}
 which is used instead of the induction equation. Since we study a steady-state accreting system, the above quantity will be constant in absence of the dynamo and the diffusion effects. This
 quantity can vary with radius due to the presence of the dynamo/diffusion effect (Machida et al. 2006; Oda et al. 2007). In the present model,
 we expect the magnetic flux advection rate varies with radius due to the presence of resistivity. We will consider this property in next
 section.
 
\section{Self-Similar Solutions}
We seek self-similar solutions in the following form (e.g. Akizuki \& Fukue 2006; Shadmehri 2008):
\begin{equation}
 \Sigma(r)=c_0 \Sigma_0 \left( \frac{r}{R} \right)^{s-\frac{1}{2}} 
\end{equation}
\begin{equation}
 v_r(r)=-c_1 \sqrt{\frac{G M}{R}} \left(  \frac{r}{R} \right)^{-\frac{1}{2}} 
\end{equation}
\begin{equation}
 \Omega(r)=c_2 \sqrt{\frac{G M}{R^3}} \left( \frac{r}{R} \right)^{-\frac{3}{2}}
\end{equation}
\begin{equation}
 c_s^2(r)=c_3 \left(\frac{G M}{R}\right) \left( \frac{r}{R} \right)^{-1}
\end{equation}
\begin{equation}
 c_A^2(r)=\frac{B_\varphi^2}{4 \pi \rho}=2 \Pi c_3 \left(\frac{G M}{R}\right) \left( \frac{r}{R} \right)^{-1} 
\end{equation}

where $\Sigma_0$ and $R$ 
are exploited to write the equations in the non-dimensional forms. Substituting the above
 solutions in the continuity, radial momentum, angular momentum, hydrostatic, and energy equations [(1),(6)-(8), and (11)], we can obtain the
 following relations: 

\begin{equation}
 c_0 c_1 =\dot{m}
\end{equation}
\begin{equation}
 -\frac{1}{2} c_1^2 + \left[ (s+\frac{1}{2})\Pi+s-\frac{3}{2} \right]  c_3 -c_2^2+1=0
\end{equation}
\begin{equation}
 -\frac{1}{2} c_0 \left[ c_1 -3 \alpha c_3 (s+\frac{1}{2}) \sqrt{1+\Pi} \right]+l^2 s \dot{m}=0
\end{equation}
\begin{equation}
 \frac{H}{r}=\sqrt{(1+\Pi) c_3}
\end{equation}
\begin{eqnarray}
 \nonumber\alpha \Pi f (\gamma-1) \left(s-\frac{1}{2}\right)^2 c_0 c_3^2+ P_m c_0 c_3 \times \\
\nonumber\left\{
\frac{9}{2} \alpha f (\gamma-1) c_2^2 
-\frac{2 c_1}{\sqrt{1+\Pi}} \left[ \left( s-\frac{3}{2} \right)\gamma - \left( s-\frac{5}{2} \right)\right]
 \right\}-\\
\frac{2\,  s\, \dot{m} \,\xi\,  P_m (\gamma-1) }{\sqrt{1+\Pi}}=0
 \end{eqnarray}
where $\dot{m}$ is non-dimensional mass accretion rate and is defined
\begin{equation}
\dot{m}=\frac{\dot{M}(R)}{2 \pi \Sigma_0 \sqrt{G M R}}.
\end{equation}

Using the equations (18)-(22), we obtain a quadratic equation for the coefficient of $c_3$:
\begin{eqnarray}
\nonumber 
-\frac{81}{16} \alpha^2 (1+\Pi) \left[ 
\frac{1+2 s}{2 l^2 s-1}
\right]^2  c_3^2 + \\
\nonumber 
\Big\{ 
\frac{3}{(\gamma-1) f} \left[ \frac{1+2 s}{2 l^2 s - 1} \right] 
\left[\left(s-\frac{3}{2}\right)\gamma-\left(s-\frac{5}{2}\right) \right]+\\
\nonumber
\frac{9}{2}\left[\left(s+\frac{1}{2}\right)\Pi+\left(s-\frac{3}{2}\right) \right]
+\frac{\Pi}{P_m} \left(s-\frac{1}{2}\right)^2
\Big\} c_3 +\\
\frac{9}{2}\left[1+\frac{2 s \xi}{3 f} \left( \frac{1+2 s}{2 l^2 s - 1} \right) \right]=0,
\end{eqnarray}
and the rest of the coefficients are
\begin{equation}
 c_0=-\frac{2}{3} \frac{\dot{m}}{ \alpha \sqrt{1+\Pi}}  \left[ \frac{2 l^2 s - 1 }{1+2 s} \right]c_3^{-1},
\end{equation}
\begin{equation}
 c_1=-\frac{3}{2}\alpha  \sqrt{1+\Pi} \left[ \frac{1+2 s}{2 l^2 s - 1} \right] c_3,
\end{equation}
\vspace{-0.3cm}
\begin{eqnarray}
 \nonumber ~~~~~~~~~~~~~~ c_2^2=1-\frac{9}{8} \alpha^2 (1+\Pi) \left[ \frac{1+2 s}{2 l^2 s - 1} \right]^2 c_3^2 
+  \\
 \left[ (s+\frac{1}{2})\Pi+s-\frac{3}{2} \right] c_3.
\end{eqnarray}
Without mass outflows, resistivity, and toroidal magnetic field, i.e. $s=l=\xi=0$, $P_m=\infty$, and $\Pi=0$, 
above relations reduce to the
 standard ADAF solutions (Narayan \& Yi 1994). Moreover, in the absence of wind and resistivity but with toroidal magnetic field,
 above relations reduce to result of Akizuki \& Fukue (2006). However the present model includes toroidal magnetic field, outflows, and resistivity.

The studies of resistive and magnetized ADAFs (e. g. Faghei 2011) imply that the solution for a set of the input parameters reaches
 to a non-rotating limit at a specific of $\Pi$ which we call it by $\Pi_c$. Assuming $c_2=0$ and
 $s=l=\xi=0$ (no wind case) in equations of (18)-(22), $\Pi_c$ can be written as
\begin{equation}
 \Pi_c=\frac{18 P_m}{f}\left[\frac{5/3-\gamma}{\gamma-1}\right].
\end{equation}
For typical values of adiabatic index and advection parameter in ADAFs, $\gamma = 4/3$ and $f = 1$, we can write $\Pi_c = 18 P_m$. 
We cannot extend the solutions for larger values of $\Pi_c$,
 because the right-hand side of equation (27) becomes negative and a negative $c_2$ is clearly unphysical. Moreover, $\Pi_c = 18 P_m$ implies
 that the flow can be magnetically dominated for $P_m > 1/18 \sim 0.056$. It means the flow can be had a strong magnetic field in the presence of
 resistivity. This property is in accord with resistive MHD simulations of Machida et al. (2006).

Now, we can investigate the radial dependence of the magnetic flux advection rate (equation 12). The self-similar solution for this
 quantity implies that 
\begin{equation}
 \dot{\Phi}(r) = \dot{\Phi}_{out}~(\frac{r}{R})^{(s-3/2)/2},
\end{equation}
 where 
$\dot{\Phi}_{out}$ is the magnetic flux advection rate at outer edge of the disc, $R$. 
Since $s < 3/2$, the magnetic flux increases with approaching to central object and the stronger wind/outflow reduces this increasing. The radial
 dependence of $\dot{\Phi}$ is qualitatively consistent with results of Machida et al. (2006) and Oda et al. (2007).

\section{Results}
Now we can investigate the behavior of the solutions in the presence of the outflows and resistivity. The effects of outflows and resistivity are 
studied via parameters $s$, $l$, $\xi$, and $P_m^{-1}$. Here, the inverse of Prandtl number specifies the resistivity of the fluid. Because, 
$\eta\propto\alpha/P_m$ and $\alpha$ parameter is $0.1$ in all Figures. The behavior of physical variables as a function of $P_m^{-1}$ are shown 
in Figures 1 and 2. The solutions in Figures 1 and 2 represent the radial inflow speed, $c_1$, and sound speed, $c_3$, both increase with the 
magnitude of resistivity ($P_m^{-1}$). These properties are qualitatively consistent with Faghei (2011). The density profiles, $c_0$, show 
that it decreases by adding resistivity. It can be due to temperature raise of the flow. The rotational velocity decreases, $c_2$, with the 
magnitude of resistivity. Because, the viscous torque increases with the temperature ($\nu \propto c_s^2 \propto T$). These properties are
 also consistent with results of Faghei (2011). 

\input{epsf}
\begin{figure} 
\begin{center}
\centerline
{ 
{\epsfxsize=4.3cm\epsffile{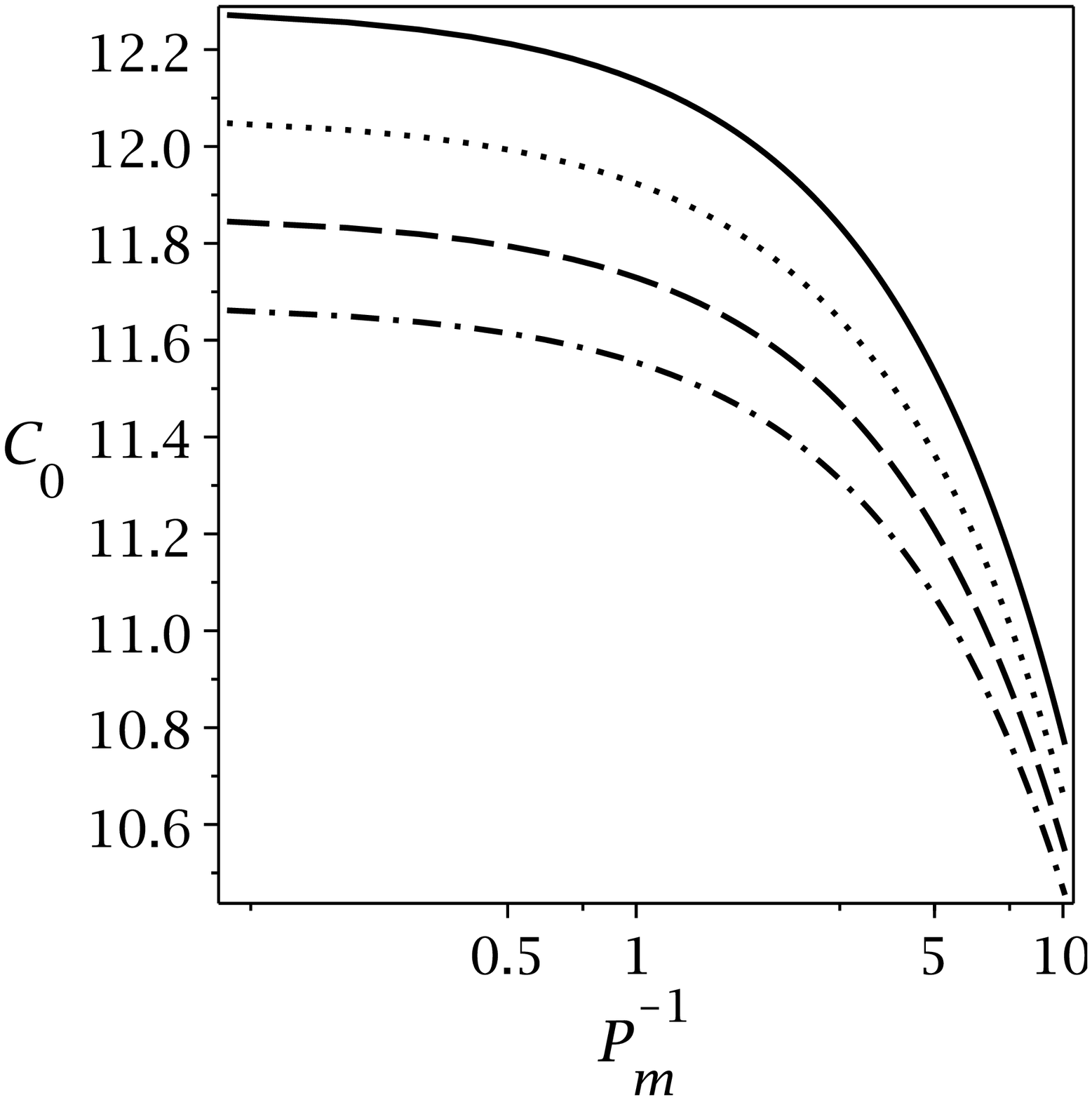} }{\epsfxsize=4.3cm\epsffile{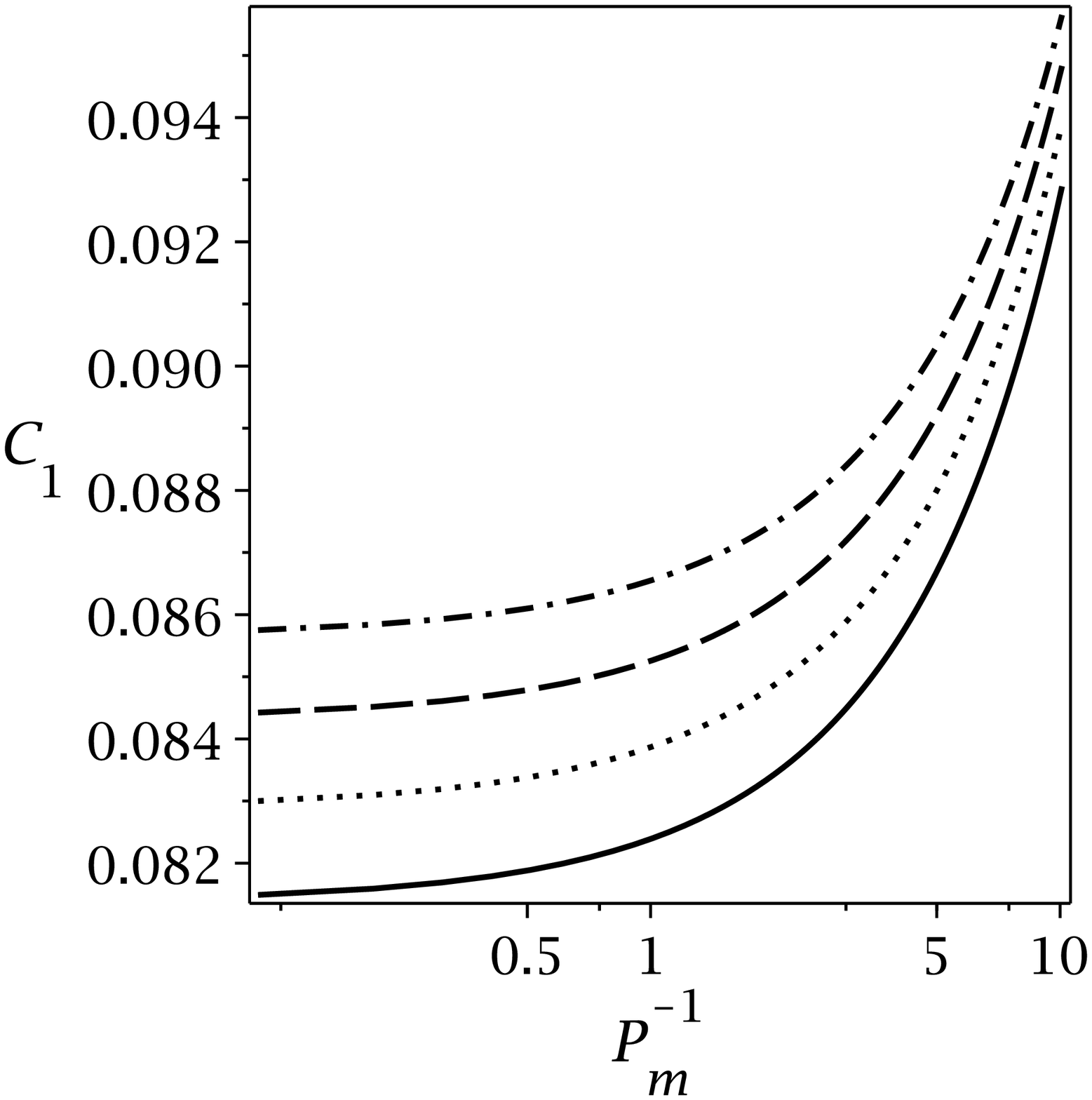}  }
} 
\centerline
{ 
{\epsfxsize=4.3cm\epsffile{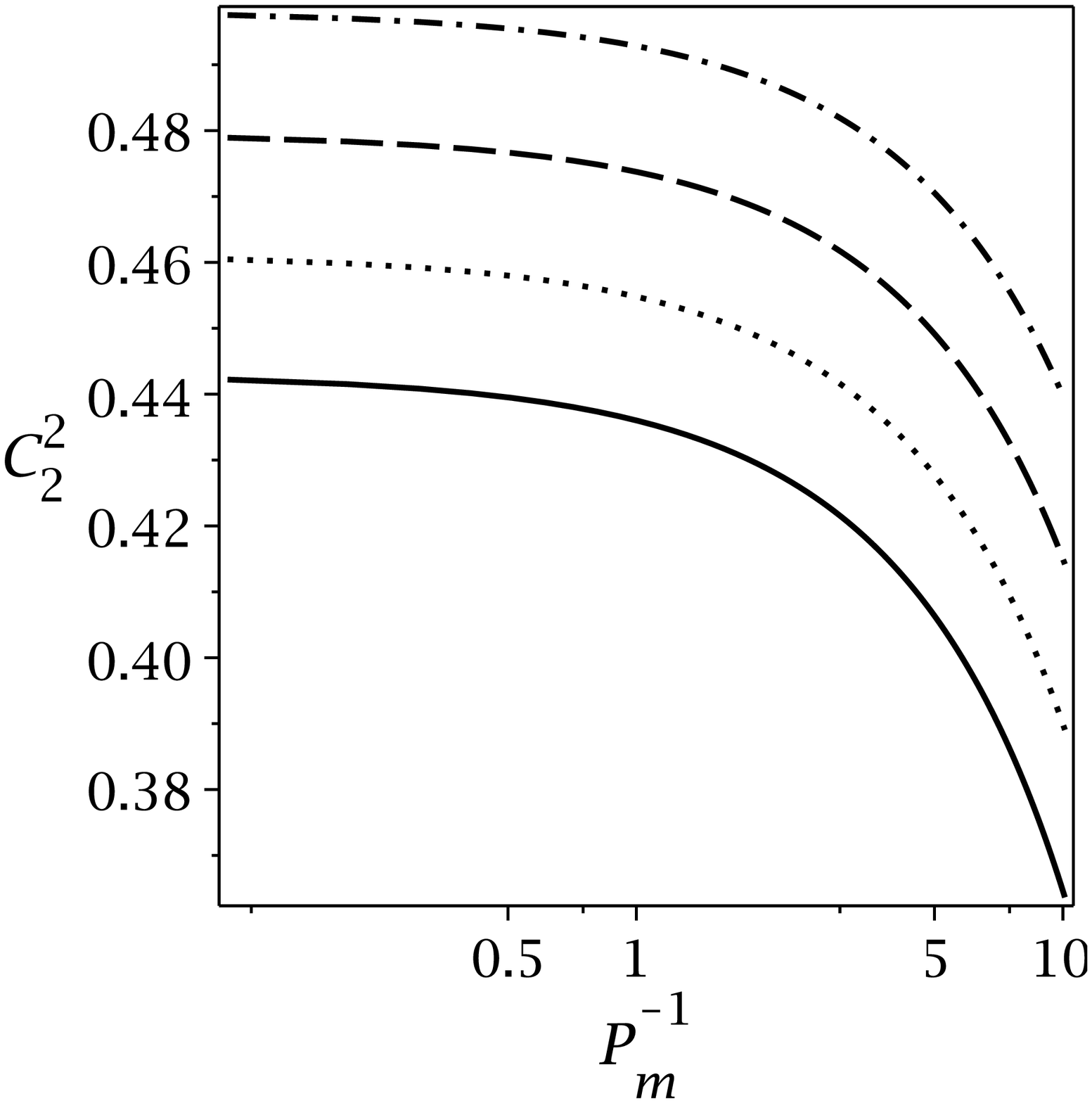}  }{\epsfxsize=4.3cm\epsffile{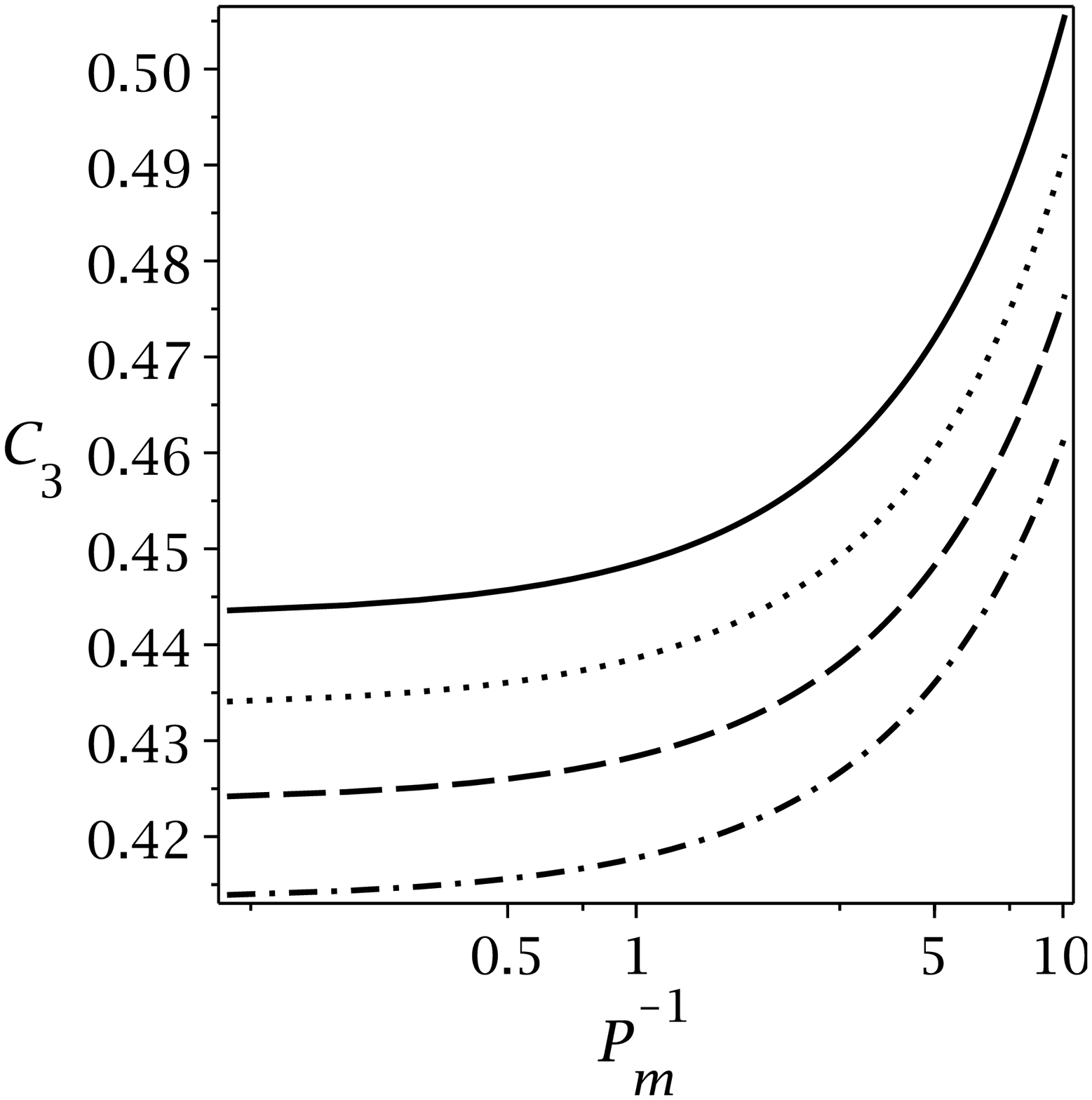}  }
}      
\end{center}

\begin{center}
\caption{Physical quantities of the flow as a function of $P_m^{-1}$ for $\gamma=4/3$, $\alpha=0.1$, $\Pi=0.5$, and  $f=l=\xi=1$. The solid, 
dotted, dashed, and dot-dashed lines represent $s=0, 0.01, 0.02$, and $0.03$. }
\end{center}
\end{figure}

In Figure 1, we also studied the effect of parameter $s$ on physical variables. The value of $s$ measures the strength of outflow and a larger $s$ 
denotes a stronger outflow. Figure 1 shows that for non-zero $s$, surface density is lower than the standard ADAF solution and for stronger 
outflows this reduction of surface density is more evidence. We can see that ADAF with wind rotates more quickly than those without winds and 
lead to enhance accretion velocity. The solution shows that temperature decreases for stronger outflows. On the other hand, outflows  play 
as a cooling agent. These properties are in accord with results of Shadmehri (2008) and Abbassi et al. (2010).

In Figure 2, the effect of energy loss due to outflows is studied by $\xi$ parameter. As with the magnitude of $\xi$ parameter, the more energy 
will carry by outflows. Due to this energy loss, we expect the temperature should decrease by adding the $\xi$ parameter.
Temperature profiles confirmed this property.  Since the turbulence viscosity is proportional to temperature ($\nu \propto T$), the efficiency 
of angular momentum transport decreases with the $\xi$ parameter. Decreasing viscous torque increases the rotational 
velocity and decreases the radial infall velocity. These properties are consistent with previous works (e. g. Shadmehri 2008). 

\input{epsf}
\begin{figure} 
\begin{center}
\centerline
{ 
{\epsfxsize=4.3cm\epsffile{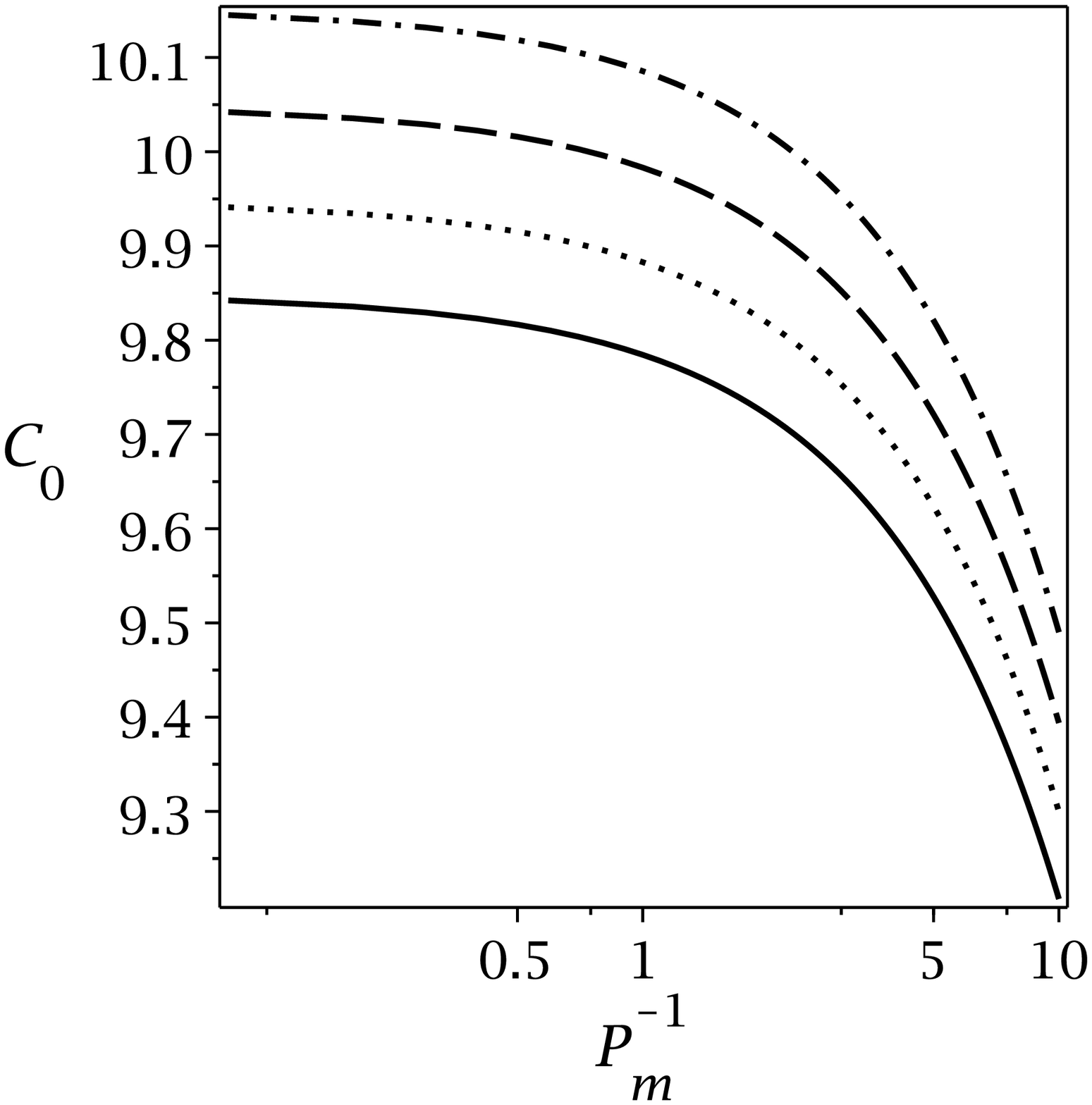}  }{\epsfxsize=4.3cm\epsffile{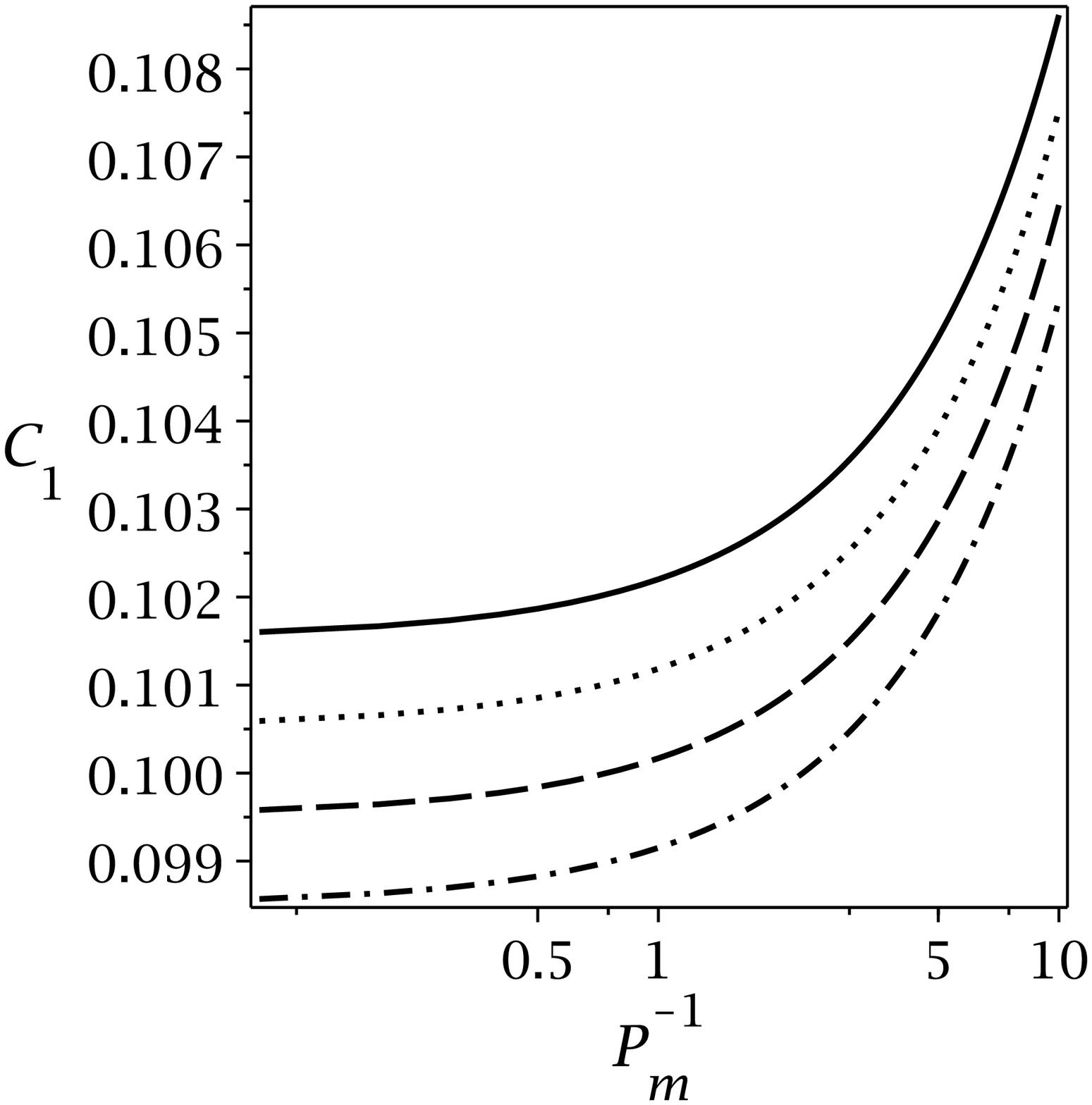}  }
} 
\centerline
{ 
{\epsfxsize=4.3cm\epsffile{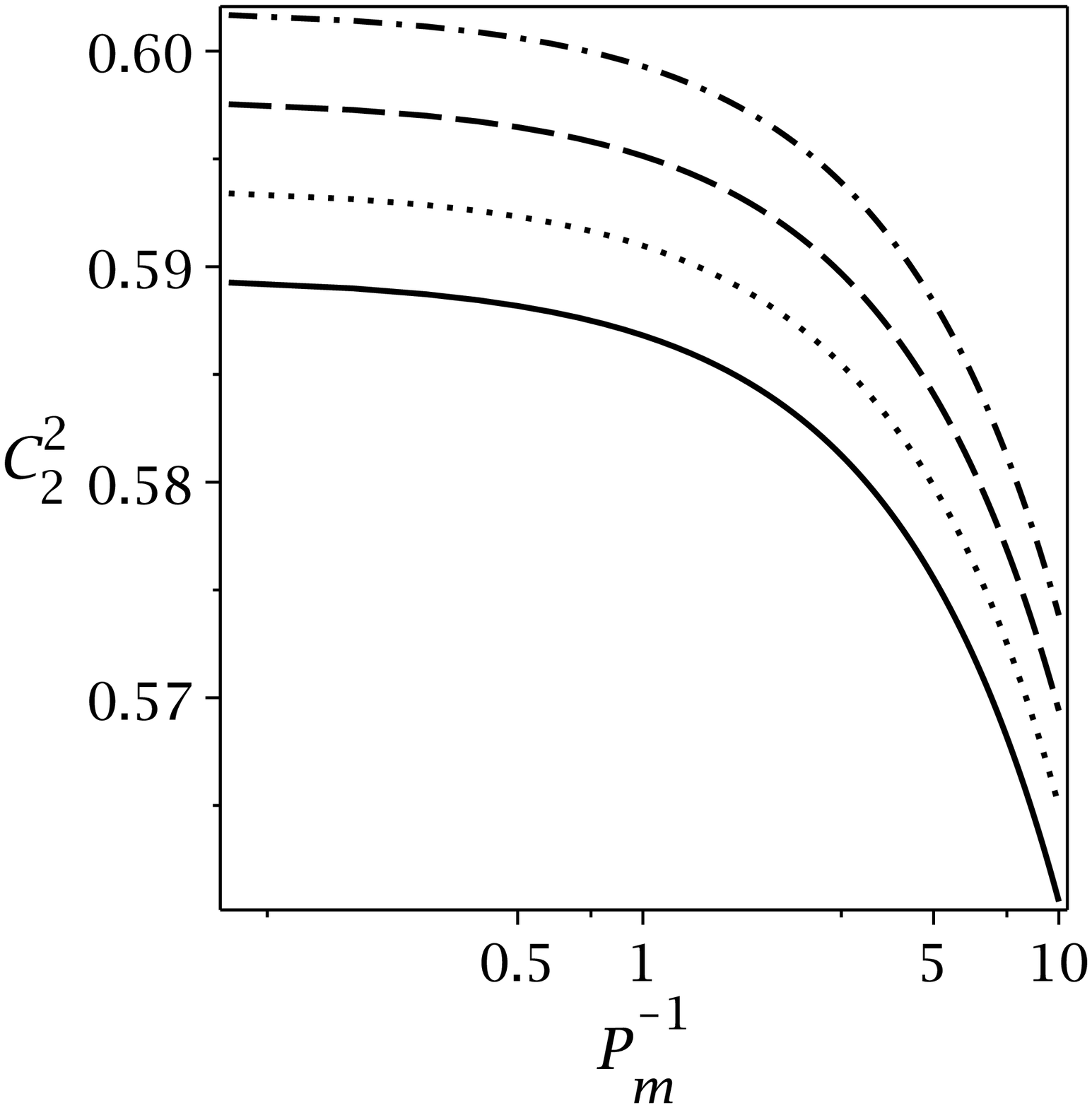}  }{\epsfxsize=4.3cm\epsffile{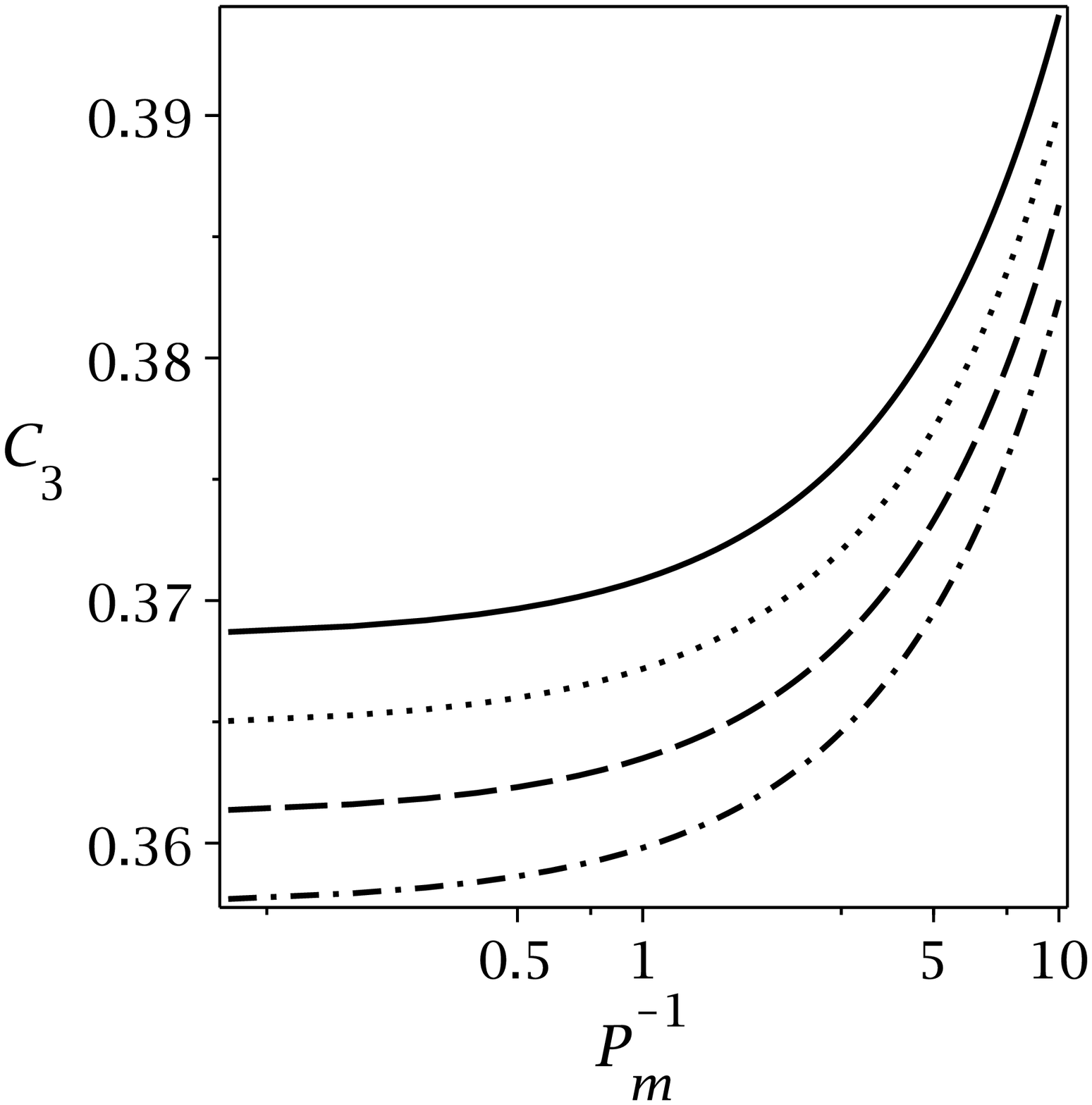}  }
}      
\end{center}

\begin{center}
\caption{Same as Figure 1, but $s=0.1$, and the solid, dotted, dashed, and dot-dashed lines represent $\xi=0, 0.1, 0.2$, and $0.3$.}
\end{center}
\end{figure}

The physical variables as a function of parameter $\Pi$ and several values of mass and energy losses are shown in Figures 3 and 4. 
By adding 
$\Pi$ which indicates the role of magnetic filed on the dynamics of accretion discs, we see the sound speed becomes larger, while surface density 
decreases. Moreover, Figure 3 represents that the radial and rotational velocities increase with the magnitude of $\Pi$. Increase in the radial 
velocity is due to the magnetic tension term dominates the magnetic pressure term in the radial momentum equation, which assists the radial 
velocity of accretion flows. Moreover, increase in the rotational velocity is because of that the disc should rotate faster than the case without 
the magnetic field which results the magnetic tension. These properties are qualitatively consistent with the previous works on magnetized ADAFs 
(e.g. Akizuki \& Fukue 2006; Khesali \& Faghei 2009; Abbassi et al. 2010). Figures 3 and 4 imply that the mass and energy losses due to wind give 
the same results by Figures 1 and 2. On the other hand, the effects of mass and energy losses on the physical variables do not change in low and 
high values of magnetic field.

Figures 5 and 6 represent that the disc thickness increases with the magnitude of the resistivity or the magnetic field. 
Because the temperature increases by adding the resistivity or the magnetic field, and from equation (21) we can see the disc thickness
increases with temperature. 
In Figures 5 and 6, the disc thickness is also studied for several values of parameters $s$ and $\xi$. The disc thickness profiles 
imply that it becomes thinner for the stronger mass or energy losses due to outflows. Because these parameters reduces the temperature of the 
flow.

\section{The Bernoulli parameter}
Here, we exploit Bernoulli parameter ($Be$) to consider the effects of the resistivity to generate/enhance outflows in magnetized ADAFs. Because, 
this parameter measures the likelihood that outflow or wind may originate spontaneously (Narayan \& Yi 1994). An adiabatic flow has a constant 
$Be$ along streamlines. If $Be$ is positive for any of accreting gas, then this gas can potentially reach infinity with a net positive kinetic 
energy. The Bernoulli parameter, defined as the sum of the kinetic energy, the enthalpy and the potential energy of the accretion flow,
\begin{equation}
 Be=\frac{1}{2}(v^2+r^2 \Omega^2)+\frac{\gamma}{\gamma-1} c_s^2-r^2 \Omega_K^2.
\end{equation}

\input{epsf}
\begin{figure} 
\begin{center}
\centerline
{ 
{\epsfxsize=4.3cm\epsffile{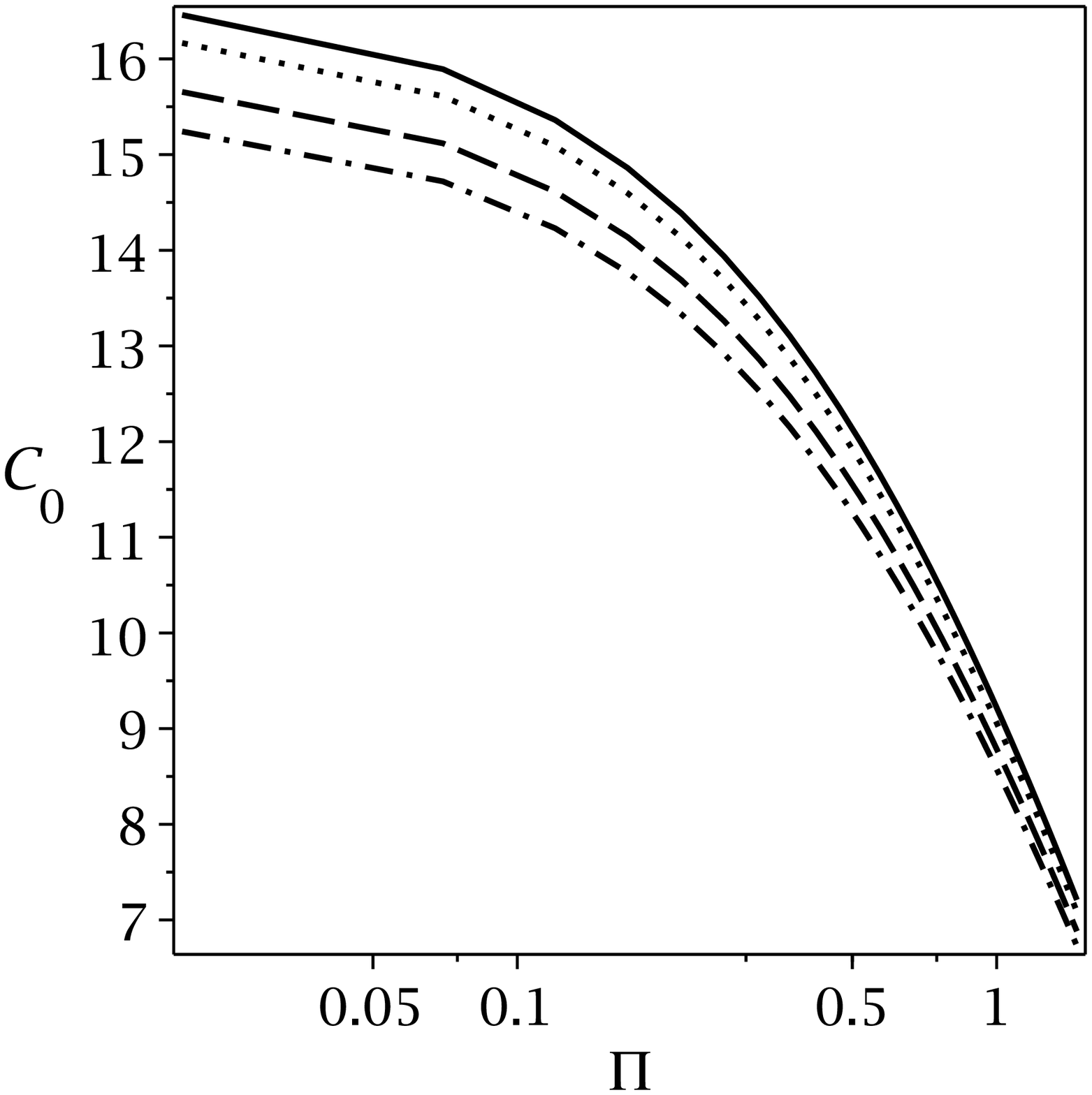}  }{\epsfxsize=4.3cm\epsffile{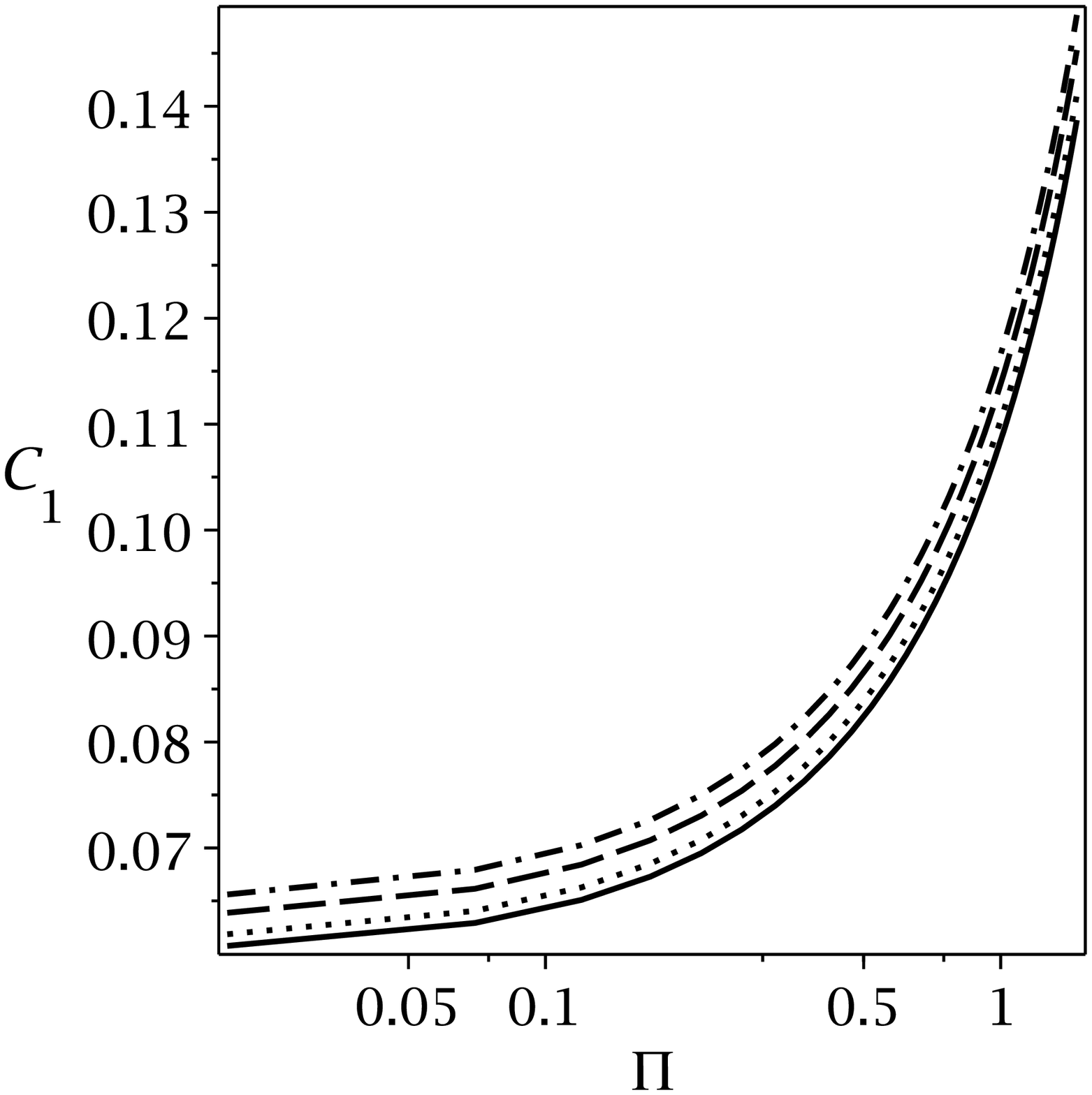}  }
} 
\centerline
{ 
{\epsfxsize=4.3cm\epsffile{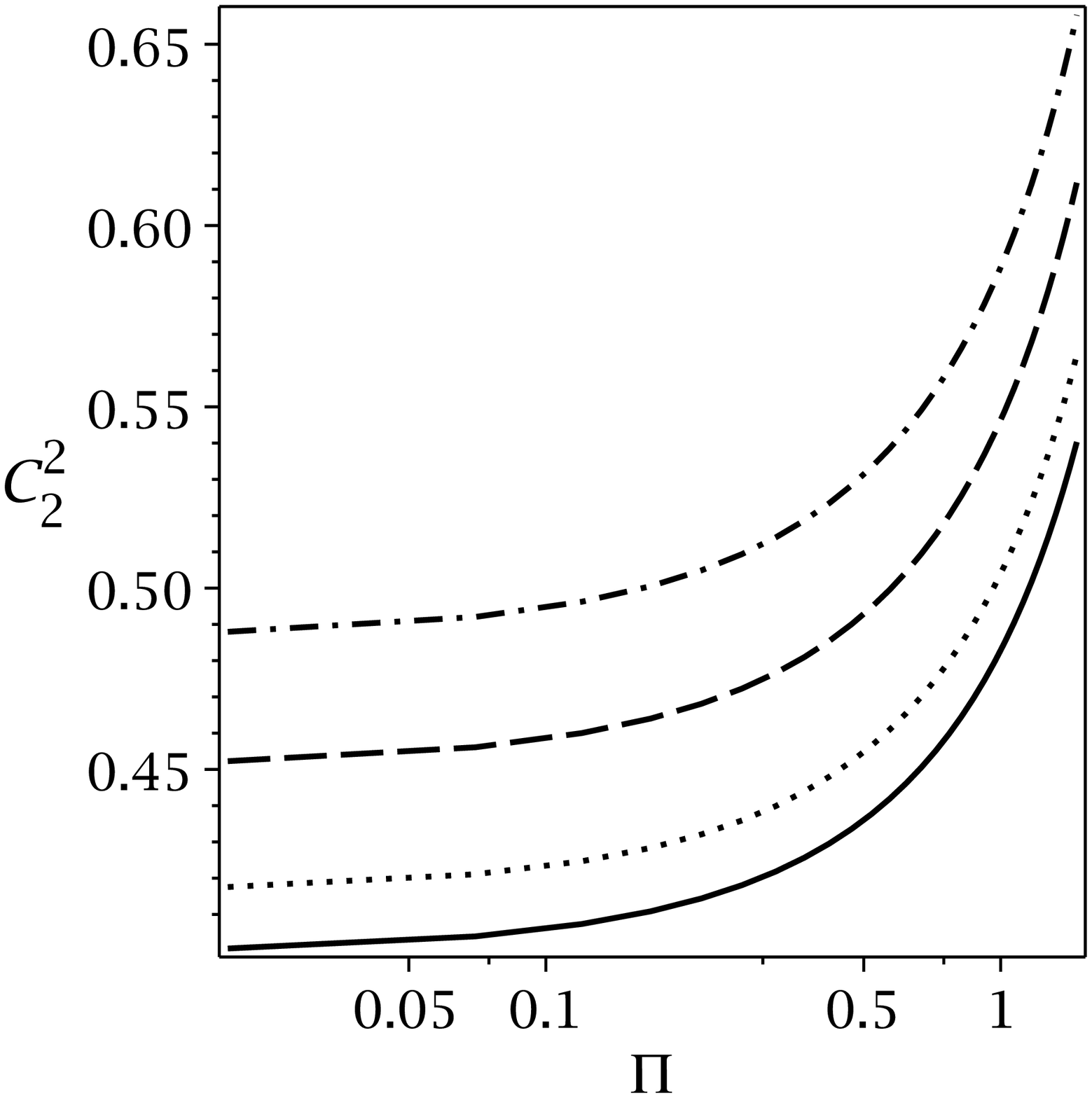}  }{\epsfxsize=4.3cm\epsffile{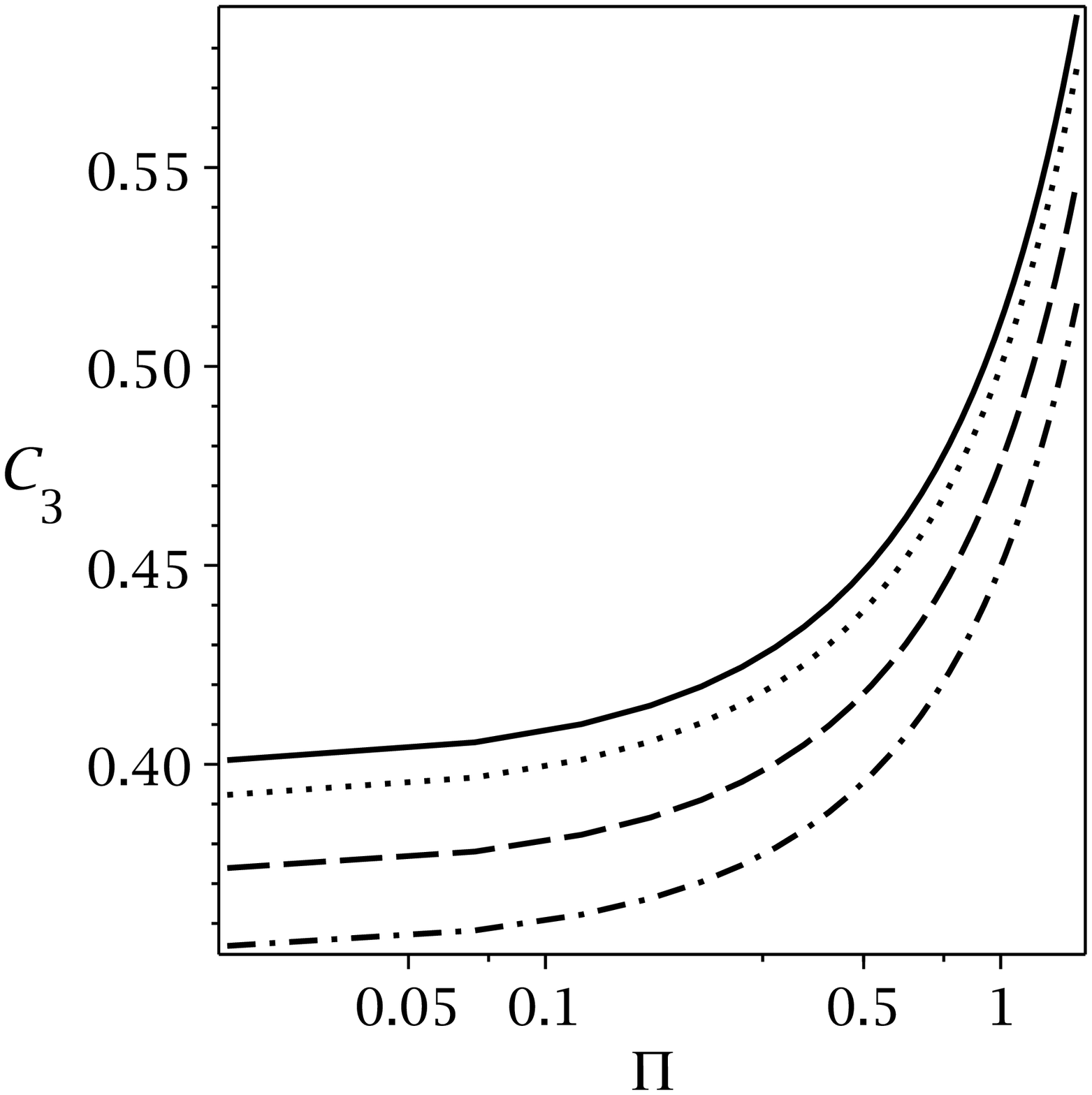}  }
}      
\end{center}

\begin{center}
\caption{Physical quantities of the flow as a function of $\Pi$ for $\gamma=4/3$, $\alpha=0.1$, $P_m=1$, $f=1$, $l=1$ and $\xi=1$. The solid, 
dotted, dashed, and dot-dashed lines represent $s=0, 0.01, 0.03$, and $0.05$.}
\end{center}
\end{figure}

Narayan \& Yi (1994) showed that Bernoulli parameter is positive in height-integrated advection dominated flows, and suggested this may explain 
the frequency occurrence of outflows and wind in many accretion systems. Using the self-similar transformations of (13)-(17), the Bernoulli 
parameter can be written as
\begin{equation}
 b=\frac{Be}{v_K^2}=\frac{1}{2}(c_1^2+c_2^2)+\frac{\gamma}{\gamma-1} c_3-1,
\end{equation}
where $b$ is the normalized Bernoulli parameter. In Figure 7, the behavior of this parameter as a function of $P_m^{-1}$ is studied for different 
values of magnetic field. The $Be$ profiles represent that it is positive and increases with the magnitude of resistivity or magnetic field. 
It can be due to increase of the flow temperature by adding the resistivity or magnetic field. Moreover, the $Be$ profiles
show the magnetic field is more important in high resistivity. Thus, the outflows can be enhanced 
in resistive and magnetized ADAFs. 

\input{epsf}
\begin{figure} 
\begin{center}
\centerline
{ 
{\epsfxsize=4.3cm\epsffile{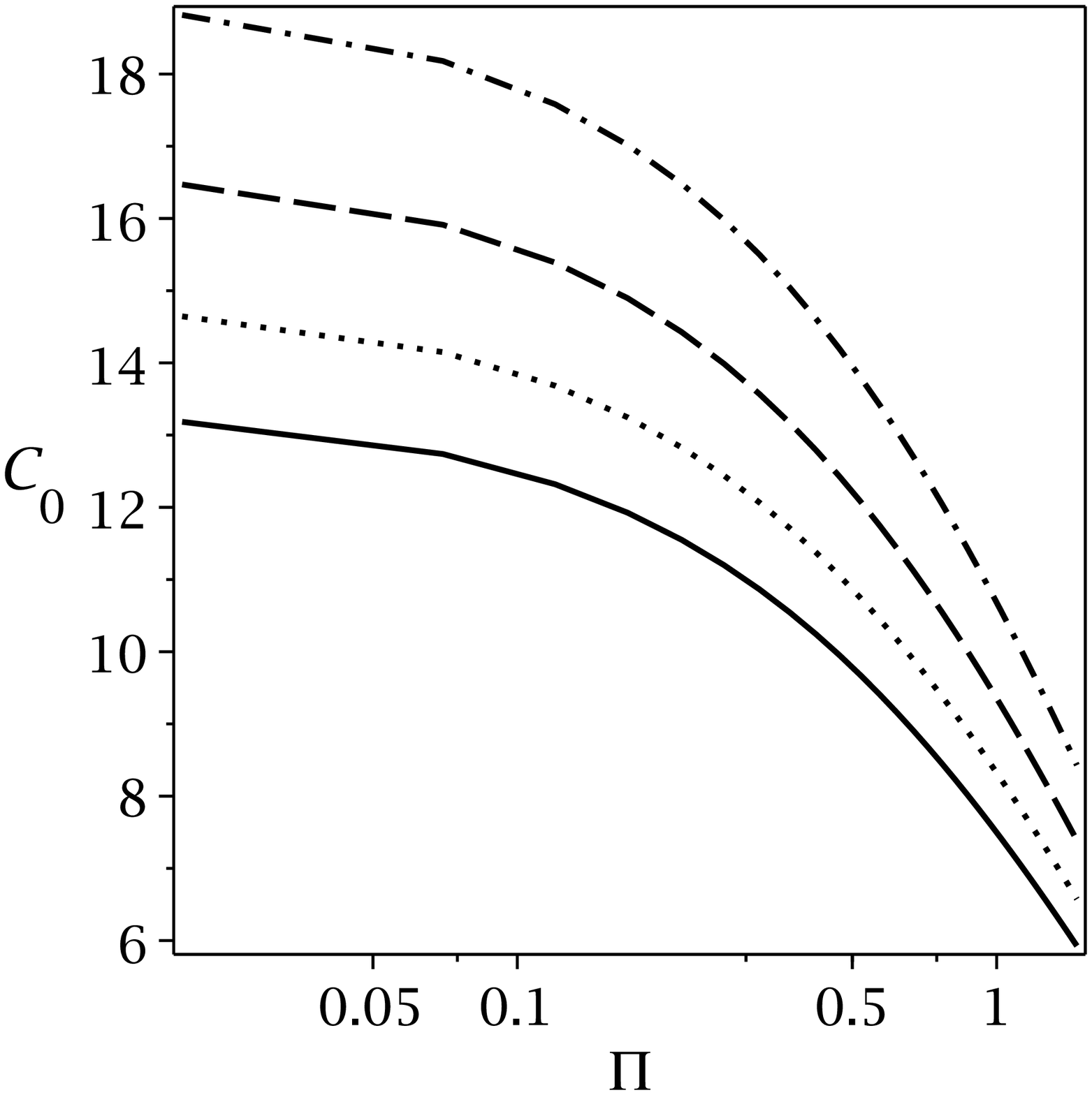}  }{\epsfxsize=4.3cm\epsffile{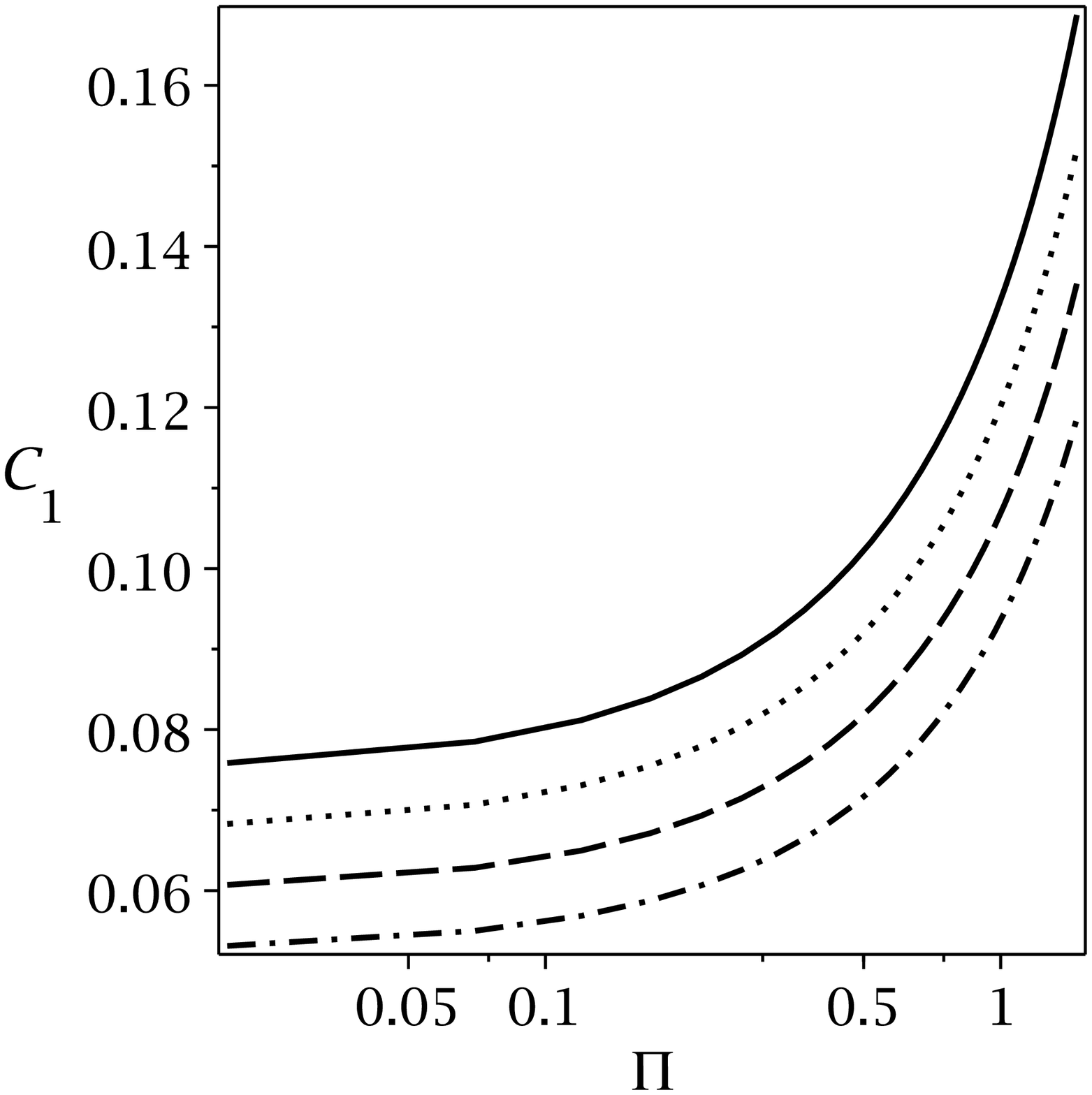}  }
} 
\centerline
{ 
{\epsfxsize=4.3cm\epsffile{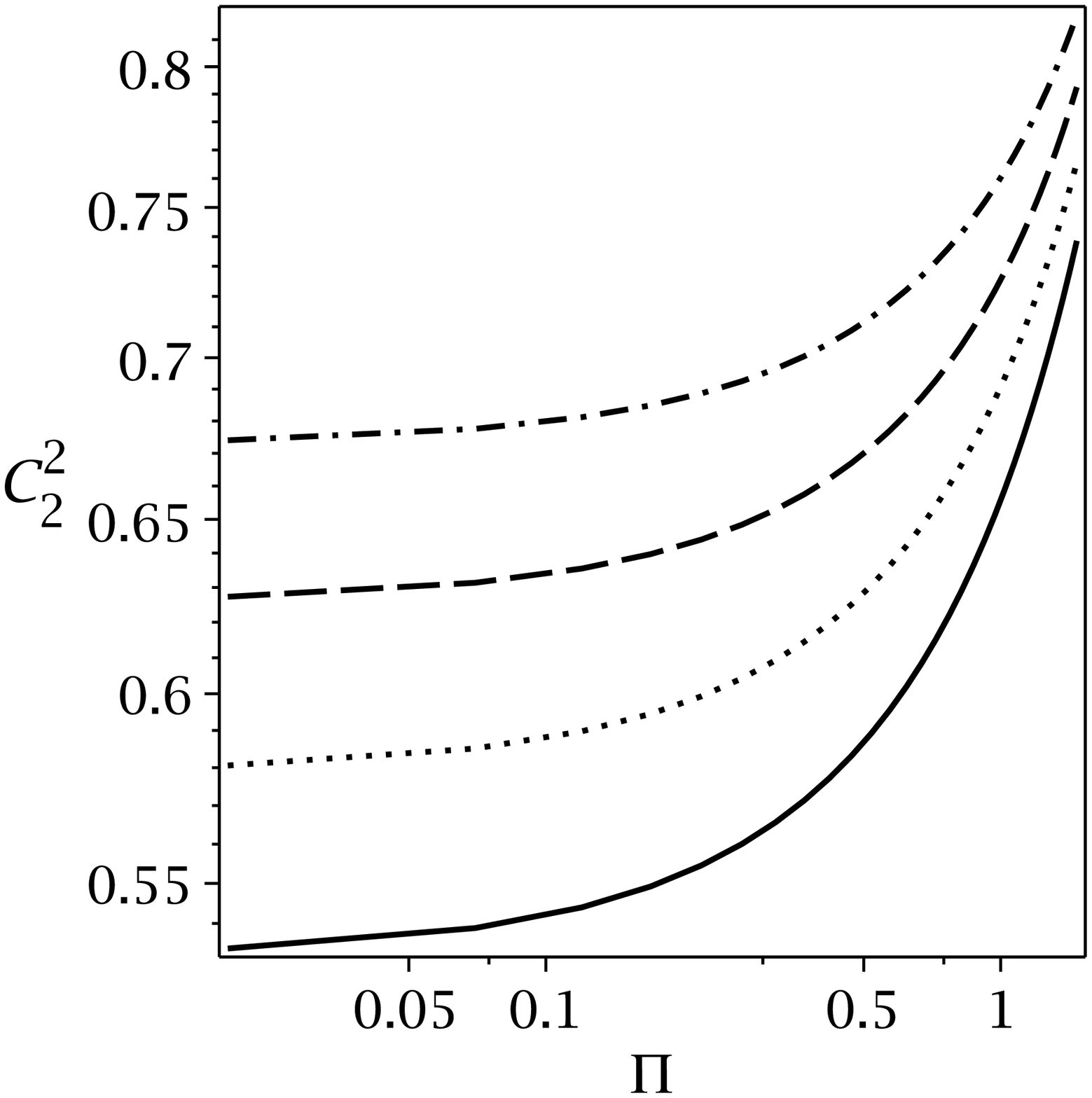}  }{\epsfxsize=4.3cm\epsffile{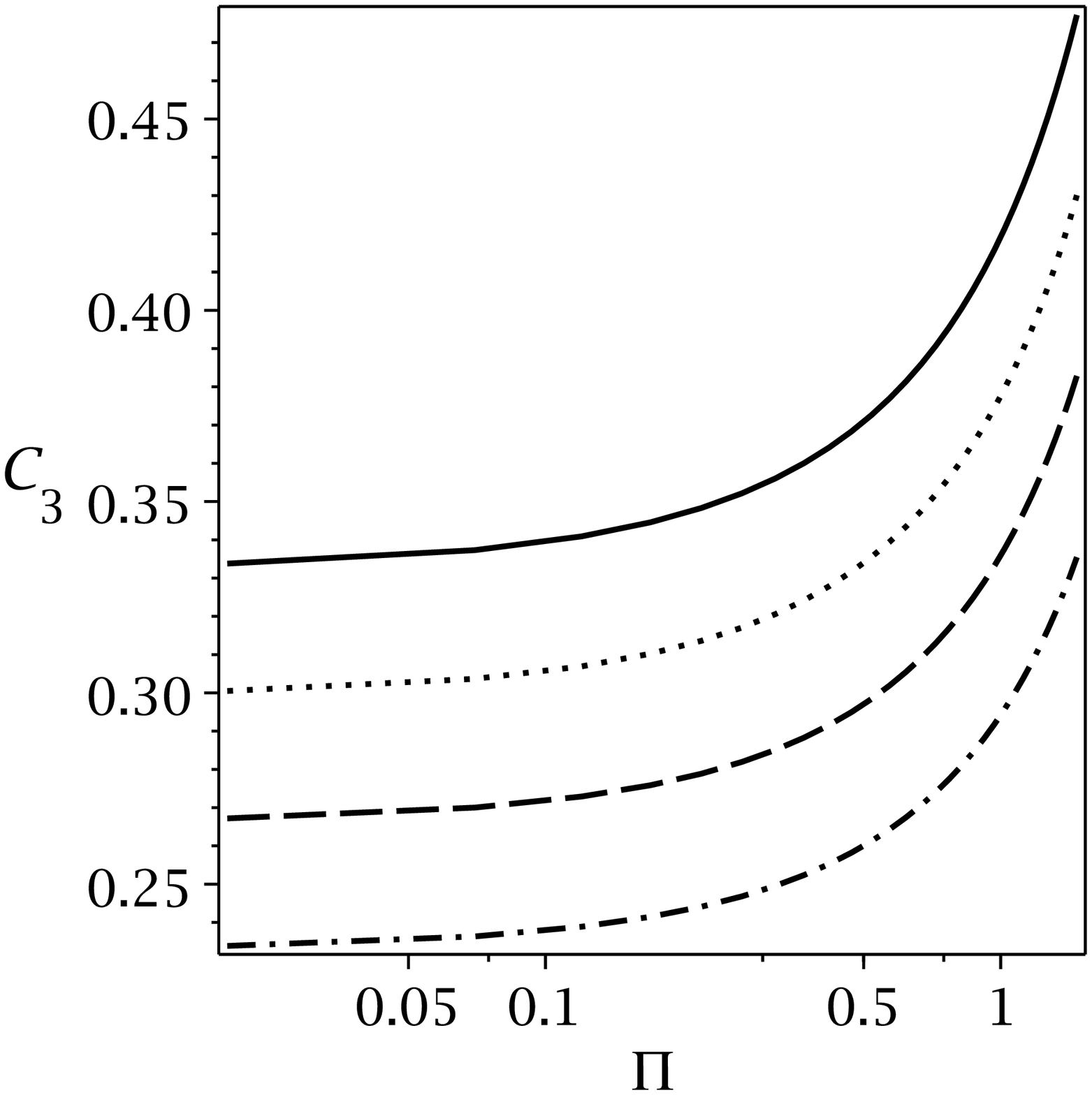}  }
}      
\end{center}

\begin{center}
\caption{Same as Figure 3, but $s=0.1$, and the solid, dotted, dashed, and dot-dashed lines represent $\xi=0, 1, 2$, and $3$.}
\end{center}
\end{figure}

\section{Summary and Discussion}
Mass loss appears to be a common phenomenon among accretion systems.
The observational evidences of ADAFs imply that outflow is important in such system. Moreover, the non-ideal MHD simulation results represent 
that resistivity can play an importance role in outflow of accretion discs. 

In this paper, the structure of a magnetized ADAF  in the presence of resistivity and outflow is investigated. We assumed that the magnetic field 
has a purely toroidal component. The outflow emanating affects on the equations of continuity, angular momentum and energy, and can therefore act 
as a sink for mass, angular momentum and energy. We adopted the presented solutions by Knigge (1999), 
Akizuki \& Fukue (2006), and Shadmehri (2008). Thus, 
we assumed that angular momentum transport is due to viscous turbulence and the $\alpha$-prescription is used for the kinematic coefficient of viscosity. 
We also assumed that the flow does not have a good cooling efficiency and so a fraction of energy accretes along with matter onto the central 
object. To solve the equations that govern the structure behavior of magnetized ADAF with outflow, we have used a steady self-similar 
solution.

\input{epsf}
\begin{figure} 
\begin{center}
\centerline
{ 
{\epsfxsize=4.3cm\epsffile{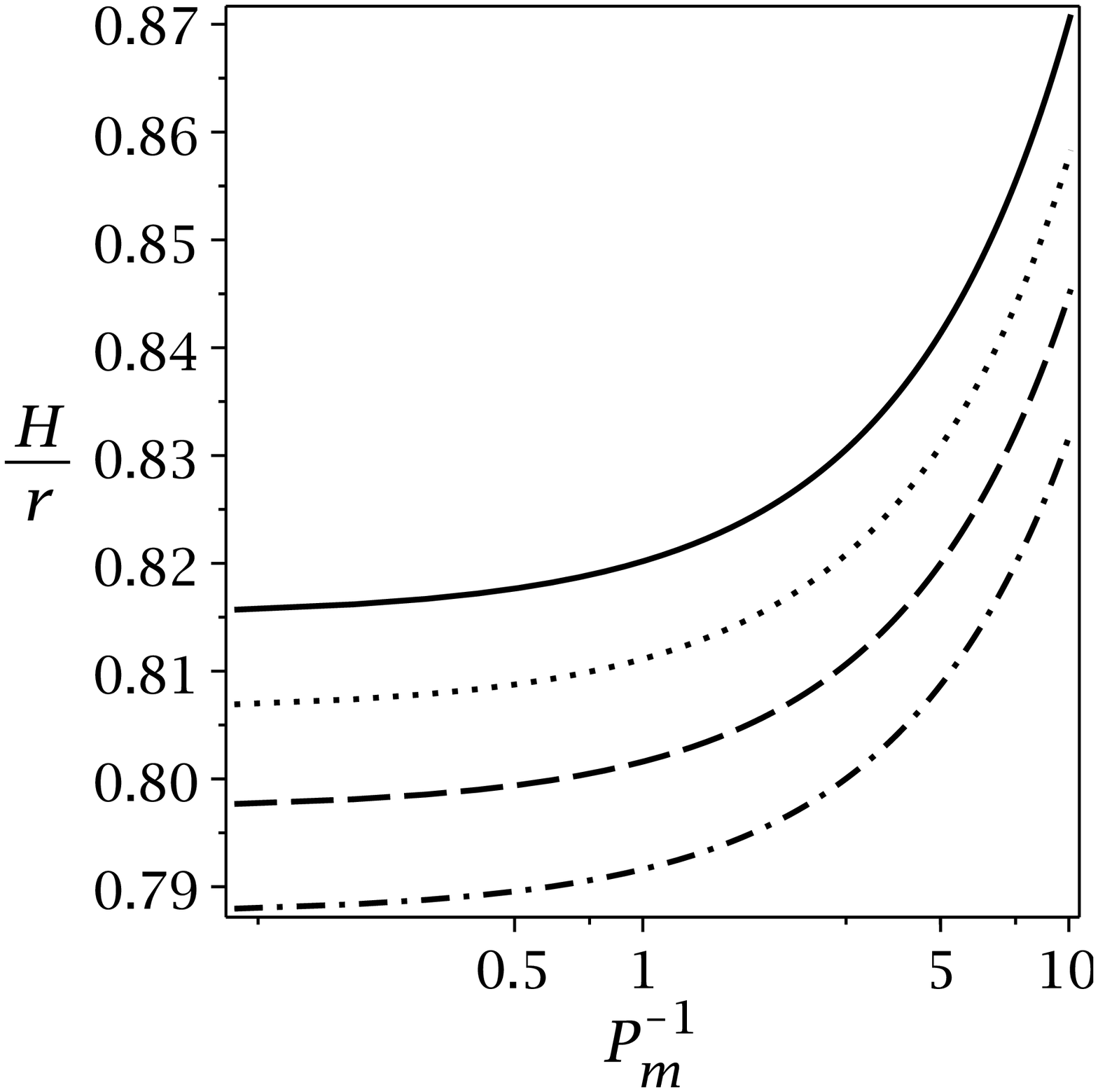}  }{\epsfxsize=4.3cm\epsffile{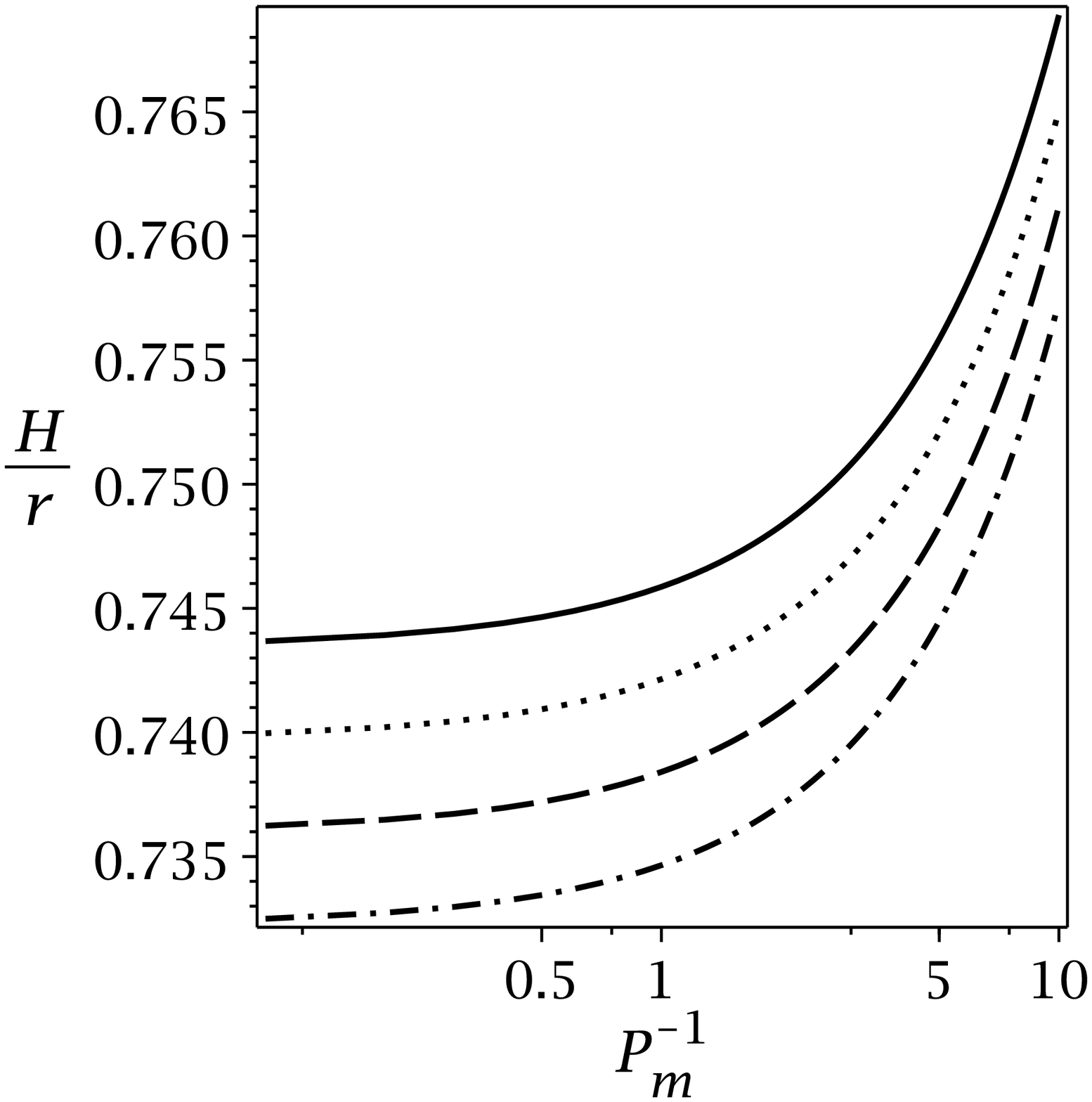}  }
} 
\end{center}

\begin{center}
\caption{The ratio of height thickness to radius as a function of $P_m^{-1}$ for different $s$ (left panel), and $\xi$ (right panel). The input 
parameters in left and right panels are same as Figures 1 and 2, respectively.}
\end{center}
\end{figure}

The present model represented the outflows in accretion flows can improve by the resistivity and magnetic field. These properties 
are in accord  with resistive MHD simulations of Fendt \& \v{C}emelji\'{c} (2002) and  \v{C}emelji\'{c} et al. (2008). For example, Fendt \& 
\v{C}emelji\'{c} (2002) showed that resistivity can affect the outflows structure in accretion discs. Moreover, they found that the outflow 
velocity increases with the magnitude of the resistivity and toroidal magnetic field. 

In this paper, we assumed a purely toroidal magnetic field that provides restrictions in the present model. 
For example, a purely toroidal magnetic 
field is not enough to  have magnetically driven outflows. Thus, the present model is unsuitable to use in disc 
with magnetically driven outflows. Moreover, we considered the presented 
model in a height-integrated approach and  applied it to calculate Bernoulli parameter. Narayan \& Yi (1995) showed that Bernoulli parameter in 
ADAF varies by latitude. As Bernoulli parameter in equatorial  is negative and becomes positive for high latitude. One can investigate 
latitudinal behavior of the present model.

\section*{Acknowledgments}
We would like to thank referee for his/her invaluable comments and very careful reading of the manuscript that helped us to
improve the initial version of the paper. 

\input{epsf}
\begin{figure}
\begin{center}
\centerline
{ 
{\epsfxsize=4.3cm\epsffile{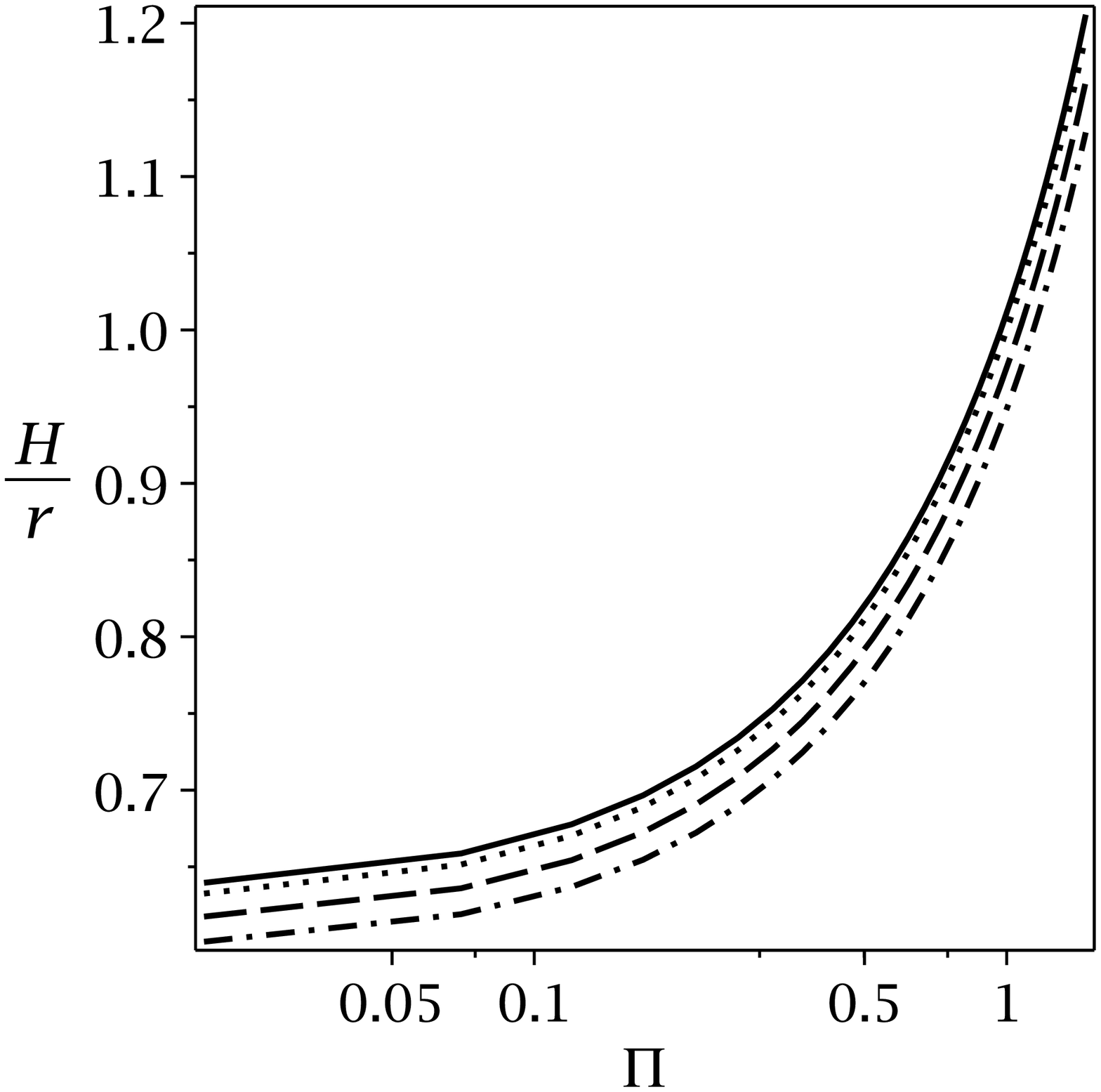}  }{\epsfxsize=4.3cm\epsffile{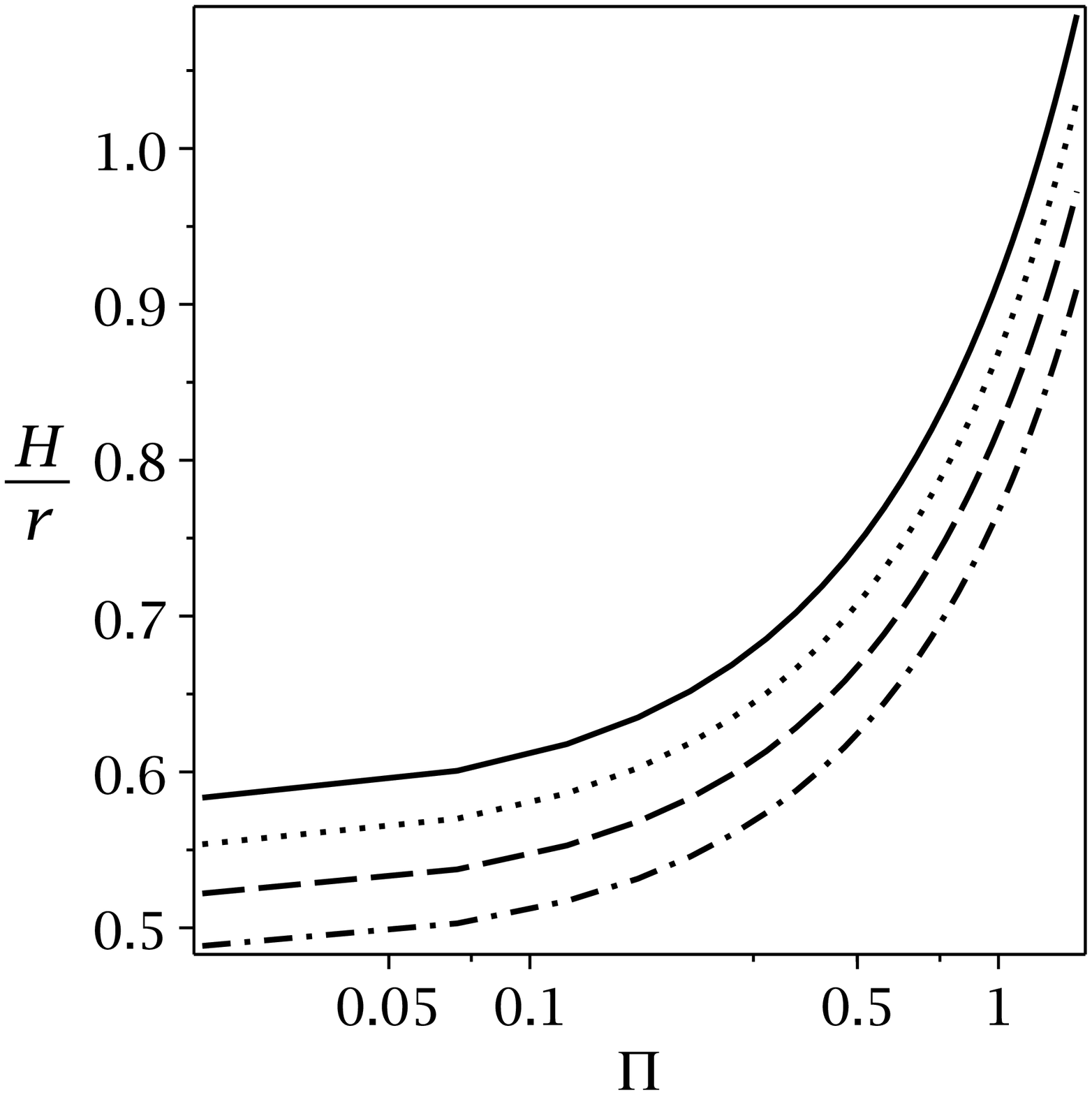}  }
} 
\end{center}

\begin{center}
\caption{The ratio of height thickness to radius as a function of $\Pi$ for different values of $s$ (left panel), and $\xi$ (right panel). The 
input parameters in left and right panels are same as Figures 3 and 4, respectively.}
\end{center}
\end{figure}

\input{epsf}
\begin{figure}
\begin{center}
\centerline
{ 
{\epsfxsize=6.0cm\epsffile{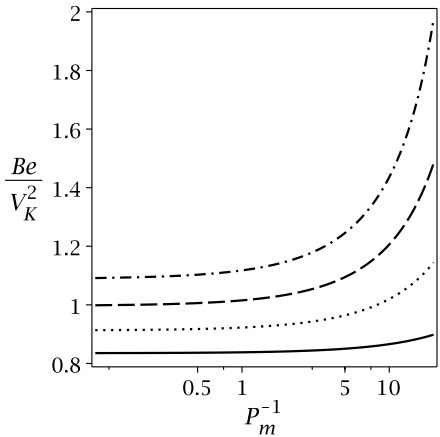}  }  
} 
\end{center}
\vspace{-1cm}
\begin{center}
\caption{The dimensionless of Bernoulli constant as a function of $P_m^{-1}$ for different values of $\Pi$. The input parameters are same as 
Figure 1, but $s=\xi=l=0$, and the solid, dotted, dashed, and dot-dashed lines represent $\Pi=0.1, 0.3, 0.5$, and $0.7$, respectively.}
\end{center}
\end{figure}


\end{document}